\documentclass[12pt,preprint]{aastex}
\usepackage{graphicx}
\begin{document}

%%%%%%%%%
% Title %
%%%%%%%%%

\title{Spectral Variability of Quasars in the Sloan Digital Sky Survey. II: The C~{\sc iv} Line\altaffilmark{1}}

%%%%%%%%%%%
% Authors %
%%%%%%%%%%%

\author{
Brian C. Wilhite\altaffilmark{2,3,4},
Daniel E. Vanden Berk\altaffilmark{5},
Robert J. Brunner\altaffilmark{3,4}, 
Jonathan V. Brinkmann\altaffilmark{6}
}

\altaffiltext{1}{Presented as part of a dissertation to the Department of Astronomy and Astrophysics, The University of Chicago, in partial fulfillment of the requirements for the Ph.D. degree.}
%\altaffiltext{2}{Based on observations obtained with the Sloan Digital
%  Sky Survey, which is owned and operated by the Astrophysical Research
%  Consortium.}
\altaffiltext{2}{The University of Chicago, Department of Astronomy
  and Astrophysics, 5640 S. Ellis Ave., Chicago, IL 60637}
\altaffiltext{3}{The University of Illinois, Department of Astronomy,
  1002 W. Green St., Urbana, IL 61801}
\altaffiltext{4}{National Center for Supercomputing Applications, 605 E. Springfield Ave., Champaign, IL 61820}
%\altaffiltext{5}{Current address: University of Illinois, Department of Astronomy, 1002 W. Green St., Urbana, IL 61801; wilhite@astro.uiuc.edu}
\altaffiltext{5}{The Pennsylvania State University, Department of Astronomy and Astrophysics, 525 Davey Lab, University Park, PA 16802}
\altaffiltext{6}{Apache Point Observatory, P.O. Box 59, Sunspot, NM 88349}

%%%%%%%%%%%%
% Abstract %
%%%%%%%%%%%%

\begin{abstract}
We examine the variability of the high-ionizaton C~\textsc{iv}$\lambda{1549}$ line in a sample of 105 quasars observed at multiple epochs by the Sloan Digital Sky Survey. 
We find a strong correlation between the change in the C~\textsc{iv} line flux and the change in the line width, but no correlations between the change in flux and changes in line center and skewness.  The relation between line flux change and line width change is consistent with a model in which a broad line base varies with greater amplitude than the line core. The objects studied here are more luminous and at higher redshift than those normally studied for variability, ranging in redshift from 1.65 to 4.00 and in absolute $r$-band magnitude from roughly $-$24 to $-$28.  Using moment analysis line-fitting techniques, we measure line fluxes, centers, widths and skewnesses for the C~\textsc{iv} line at two epochs for each object.  The well-known Baldwin Effect is seen for these objects, with a slope $\beta = -0.22$.  The sample has a median intrinsic Baldwin Effect slope of $\beta_{int} = -0.85$; the C~\textsc{iv} lines in these high-luminosity quasars appear to be less responsive to continuum variations than those in lower luminosity AGN.  Additionally, we find no evidence for variability of the well known blueshift of the C~\textsc{iv} line with respect to the low-ionization Mg~\textsc{ii}$\lambda$2798 line in the highest flux objects, indicating that this blueshift might be useful as a measure of orientation.  
\end{abstract}


%%%%%%%%%%%%%%%%
% Introduction %
%%%%%%%%%%%%%%%%

\section{Introduction\label{intro}}

Quasar emission lines represent light reprocessed by high-velocity ionized gas surrounding a central continuum source.  As the central source continuum varies, the emission lines vary in response.  Quasar continuum variability is well studied.  Long-known anticorrelations between variability amplitude and luminosity \citep[e.g.,][]{uomoto76,cristiani97} and between variability amplitude and wavelength \citep[e.g.,][]{giveon99,trevese01,wilhite05}, as well as the correlation between variability and time lag \citep[e.g.,][]{hook94,hawkins02,devries03} were recently parameterized by \citet{vandenberk04}.

Variability of quasar emission lines is also well studied, but for a much smaller number of objects.  Much of this work has come as a part of reverberation mapping efforts to determine black hole masses and study the structure of the broad-line region \citep[e.g.,][]{peterson93,wandel99}, and has generally focused on line response to continuum variability.  These emission line variability studies have been critical in characterize the structure of the broad-line region.  By measuring the response delays of various emission lines, with respect to the continuum, reverberation mapping has shown that the broad-line emitting region species are stratified by ionization potential \citep[e.g.][]{peterson93}, and that the size of the BLR is dependent upon continuum luminosity \citep[e.g.][]{wandel99,kaspi05}.

Until recently, C~\textsc{iv}$\lambda{1549}$ had been monitored in only a few low-redshift, low-luminosity objects, such as NGC5548, monitored for years with both IUE and HST \citep{clavel91,korista95}, and NGC 5141, observed for short-term variability with IUE \citep{crenshaw96}.  \citet{kaspi03} have begun a campaign to use C~\textsc{iv} variability in reverberation mapping of high-redshift, high-luminosity quasars, but the results are not yet conclusive.

Profile variability of AGN emission lines has been studied \citep[e.g.,][]{wanders96,sergeev01}, but most of this has been done for nearby, low-luminosity Seyferts and has been limited to the rest-frame optical.  The reasons for this are well motivated---low-luminosity objects are known to be more variable and optical spectroscopy is more common---but this means there has been relatively little study of line variability in higher luminosity objects, or of rest-frame ultraviolet lines like C~\textsc{iv}.

Although not well studied in the time domain, the C~\textsc{iv} line has demonstrated several intriguing properties at a single epoch which suggest study of C~\textsc{iv} variability could prove useful in understanding the structure of the broad emission line region and quasars as a whole.

\subsection{The C~\sc{iv} Line Profile\label{introprofile}}

\citet{wills92} and \citet{francis92} found that C~\textsc{iv} line width is anticorrelated with the equivalent width of the line, a result even more clearly demonstrated by \citet{wills93}.  The highest flux C~\textsc{iv} lines tend to be the most narrow.  \citet{wills93} suggested that this might be due to different relative importances of an intermediate width line region (ILR) and the very broad line region (VLBR) from quasar to quasar.  In this scenario, the narrow ($\sim2000$ km/s) line core is produced in the ILR, which lies near the outer edge of the broad-line region (BLR).  The broad ($\sim7000$ km/s) line base is a product of the VBLR, which comprises the inner portion of the BLR.  According to the ILR model, possible line core fluxes extend over a larger range in values than the line base fluxes.  Thus, the width of an individual line is strongly dependent upon the strength of the ILR line core.  Strong C~\textsc{iv} lines have dominant cores and are therefore narrow.  Similarly, weak C~\textsc{iv} lines have less dominant cores and, thus, relatively more important line bases, and are preferentially broader as a result.  \citet{murray97} found that the C~\textsc{iv} profile could be reproduced with a continuous disk-wind model and did not distinguish between intermediate-width and very broad line regions.

\citet{wills93} also found that the C~\textsc{iv} line is typically asymmetric, in the sense that C~\textsc{iv} lines are generally skewed to shorter wavelengths.  \citet{richards02a} found that this asymmetry tends to increase with increasing C~\textsc{iv} blueshift; the most blueshifted lines also tend to be the most skewed toward the blue end of the spectrum.

\subsection{The Baldwin Effect in the C~\sc{iv} Line\label{introbaldwin}}

\citet{baldwin77}, and others later \citep{kinney90,baskin04} demonstrated that the equivalent width of the C~\textsc{iv} line is anticorrelated with the luminosity of the nearby continuum for quasars observed at a single epoch:

\begin{equation}
W_{C~\textsc{iv}} \propto L_{\lambda}^{\beta}
\end{equation}

The initial fit to the slope was $\beta=-0.64$ \citep{baldwin78}.  \citet{kinney90} found a lower value of $\beta = -0.17 \pm 0.04$ and later studies \citep[c.f.,][]{dietrich02} have found similar results.

The Baldwin Effect may also be recast in terms of line luminosity, giving the similar 

\begin{equation}
L_{C~\textsc{iv}} \propto L_{\lambda}^b,
\end{equation}
where $b=\beta + 1$. In this form, the relation is easier to understand.  From quasar to quasar, as the continuum luminosity 
increases, the C~\textsc{iv} line luminosity increases, but at a slower rate.  By using equivalent width as a proxy for luminosity,   it had originally been hoped that the Baldwin Effect could be used as a cosmological probe.  Unfortunately, the roughly half-magnitude scatter about the original relation is too large to allow for precision cosmology.

\citet{richards02a} found that the Baldwin Effect in C~\textsc{iv} appears to be related to the blueshift 
of the line with respect to lower ionization lines, such as Mg~\textsc{ii} (see \S\,\ref{introlineshifts}).  
They separated almost 800 quasars into four equally populated bins, splitting by the size of 
the blueshift; from the quasars in each bin, they created a composite spectrum.  They found a 
clear anticorrelation between the strength of the C~\textsc{iv} line and the size of the blueshift.  The bin with 
the largest C~\textsc{iv}-Mg~\textsc{ii} blueshift also had the lowest equivalent width composite C~\textsc{iv} line, and vice versa.  

%Of equal interest, lines 
%which demonstrate no Baldwin Effect, such as Si~\textsc{iv}, also demonstrated no differences in profile 
%between bins.  Only blueshifted lines showed signs of a Baldwin Effect. 

It has been suggested that the C~\textsc{iv} Baldwin Effect could be largely reproduced through a 
softening of the continuum slope with increasing luminosity and luminosity-dependent quasar 
metallicity \citep{korista98}.  \citet{wang98} found that the ultraviolet to X-ray spectral 
index is correlated with quasar luminosity: more luminous quasars have softer ionizing continua 
slopes.  They also found that the UV to X-ray index is strongly correlated with C~\textsc{iv} equivalent 
width.  The combination of these effects leads directly to the Baldwin Effect: quasars with 
high luminosity display low C~\textsc{iv} equivalent width.  Though these relations are consistent with 
the Baldwin Effect, the physical driver itself is not yet understood.

%That some high-ionization lines (such as 
%NV or SiIV) show no signs of a Baldwin Effect may be explained by a correlation of metallicity 
%and luminosity.  As luminosity increases, the increased importance of the secondary cooler 
%NV balances the decrease in NV flux due to the softening of the spectral slope.  CIV line 
%strength is not strongly dependent on metallicity because of its role as a strong cooling line.

Recently, \citet{baskin04} found that the correlation between C~\textsc{iv} equivalent width was 
much stronger with $L^{1/2}(H_{\beta} FWHM)^{-2}$, a proxy for $L/L_{EDD}$ (since the black hole 
mass scales as $L^{1/2}(H_{\beta} FWHM)^{2}$), than it was with the simple continuum luminosity.
They have suggested that the Baldwin Effect may in fact be a secondary effect spawned by a more 
fundamental relation between C~\textsc{iv} equivalent width and the relative accretion rate, $L/L_{EDD}$.  
However, the potential physical mechanism driving the relation is unknown.

The roughly half-magnitude scatter in the Baldwin Effect was shown by \citet{kinney90} to be at least partially due to continuum and C~\textsc{iv} line variability.  
As a quasar's continuum luminosity increases or decreases, the C~\textsc{iv} line luminosity (which consists largely of reprocessed continuum photons) increases or decreases in turn, with a small delay owing to the light travel time.  An intrinsic relationship between continuum and line luminosities may be written in forms identical to those for the global Baldwin relation ($W_{C~\textsc{iv}} \propto L_{\lambda}^{\beta_{int}}$ and $L_{C~\textsc{iv}} \propto L_{\lambda}^{b_{int}}$).  \citet{kinney90} found that the so-called ''intrinsic Baldwin Effect'' (IBE) slope ranged from $\beta$ = -0.4 to -0.9 for 6 Seyfert galaxies and 3C 273 with an average of $\beta_{int} \approx -0.65$ ($b_{int} \approx 0.35$).  

The intrinsic Baldwin Effect is, for historical reasons, usually cast in terms of equivalent 
width, but is more straightforward when expressed in terms of luminosity.  As was the case with 
the global Baldwin Effect, the slope of the IBE is between 0 and 1.  This indicates that, for 
an individual quasar, the BLR reprocessing of the incident continuum light is not perfectly 
efficient.  As the continuum luminosity of an individual quasar fluctuates, so too does the 
CIV line luminosity, but to a lesser degree.  If an object had an intrinsic Baldwin Effect slope of 0.35, a doubling in a quasar's continuum luminosity would only lead to a roughly 25\% increase 
in CIV line luminosity (after allowing for the light-travel time delay).

The intrinsic Baldwin Effect (IBE) slope itself has been found to vary.  Over 13 years of monitoring, 
the IBE slope of the H$_{\beta}$ line in the Seyfert I galaxy NGC 5548 ranged from $b=0.4$ to 1.0 
on time scales of roughly one year \citep{goad04}.  The slope was strongly anti-correlated with 
continuum flux, indicating a lower line responsivity at higher continuum flux levels, which is 
consistent with photoionization models \citep{korista04}.

\citet{pogge92} determined that the Baldwin Effect scatter may be further reduced by accounting for the light travel time, $\tau$, between the continuum source and the broad emission line region: $L_{C~\textsc{iv}}(t) \propto L(t-\tau)^{\beta}$.  

\subsection{C~\sc{iv} Line Shifts\label{introlineshifts}}

\citet{gaskell82} first demonstrated that high-ionization quasar broad emission lines (such as C~\textsc{iv}) are typically blueshifted by hundreds of kilometers per second with respect to the low-ionization lines, thought to represent the true systemic redshift of the quasar.  This was verified in a number of later studies \citep{wilkes84,espey89,corbin90,tytler92,mcintosh99,sulentic00}.  Recently, \citet{richards02a} measured the blueshift of the C~\textsc{iv} line with respect to Mg~\textsc{ii} for $\sim800$ quasars in the SDSS Early Data Release Quasar Catalog \citep{schneider02}.  A possible correlation between C~\textsc{iv} blueshift and radio-determined orientation measures, as well as a similarity between the spectra of broad absorption line quasars and quasars with large C~\textsc{iv} blueshifts, prompted \citet{richards02a} to suggest the possibility that C~\textsc{iv} blueshift could be used as a measure of quasar orientation, either internal (related to the disk wind opening angle) or external (related to the line of sight to the observer).
They proposed that the blueshift might be a result of the obscuration or suppression of the C~\textsc{iv} flux on the red side of the line.  If the blueshift of the C~\textsc{iv} line, relative to low-ionization lines like Mg~\textsc{ii}, is related to the observer's viewing angle, it could represent the first technique to measure orientation for radio-quiet quasars.  

\subsection{The Present Work\label{presentwork}}

This is the second paper reporting results of a quasar spectral variability program using
data from the Sloan Digital Sky Survey \citep[SDSS;][]{york00}.  The first paper \citep[][hereafter Paper I]{wilhite05} examined the detailed wavelength dependence of quasar variability.  This paper focuses on the high-ionization C~\textsc{iv}$\lambda1549$ line.

We briefly summarize the SDSS data acquisition, our previous spectrophotometric re-calibration work, and the creation of the variable quasar sample in \S\,\ref{dataset}.  In \S\,\ref{linefitting}, we describe the line-fitting algorithm used here.  The variability of the C~\textsc{iv} line flux and profile is studied in \S\,\ref{civvariability}.  Interesting individual objects are identified in \S\,\ref{individual}.  The results are discussed in \S\,\ref{discussion} and we conclude in \S\,\ref{conclusions}.

Throughout the paper we assume a flat, cosmological-constant-dominated cosmology 
with parameter values $\Omega_\Lambda = 0.7, \Omega_{M} = 0.3,$ and $H_{0}=70$km s$^{-1}$ Mpc$^{-1}$.

%%%%%%%%%%%%%%%
% The Dataset %
%%%%%%%%%%%%%%%

\section{The Sloan Digital Sky Survey and the Variable Quasar Sample\label{dataset}}
\subsection{The Sloan Digital Sky Survey\label{SDSS}}

Through Summer 2004, the Sloan Digital Sky Survey \citep{york00} had imaged almost $\sim8200$deg$^{2}$ and obtained follow-up spectra for roughly $5 \times 10^{5}$ galaxies and $5 \times 10^{4}$ quasars.  All imaging and spectroscopic observations are made with a dedicated 2.5-meter telescope at the Apache Point Observatory in the Sacramento Mountains of New Mexico.  Imaging data are acquired by a 54-chip drift-scan camera \citep{gunn98} equipped with the SDSS $u,g,r,i$ and $z$ filters \citep{fukugita96}; they are then reduced and calibrated by the PHOTO software pipeline \citep{lupton01}.   The photometric system is normalized such that SDSS magnitudes are on the AB system \citep{smith02}.  A 0.5-meter telescope monitors site photometricity and extinction \citep{hogg01}.   
Point source astrometry for the survey is accurate to less than $100$ milliarcseconds \citep{pier03}.  \citet{ivezic04} discusses imaging quality control.

Objects are targeted for follow-up spectroscopy as candidate galaxies \citep{strauss02,eisenstein01}, quasars \citep{richards02b} or stars \citep{stoughton02}.  Targeted objects are grouped in 3-degree diameter "tiles" \citep{blanton03} and aluminum plates are drilled with 640 holes whose locations on the plate  correspond to the objects' sky locations.  Each plate is placed in the imaging plane of the telescope and plugged with optical fibers assigned to roughly 500 galaxies, 50 quasars and 50 stars.  Fibers run from the telescope to twin spectrographs.

SDSS spectra cover the observer-frame optical and near infrared, from 3900\AA---9100\AA.
Spectra are obtained in three or four consecutive 15-minute observations until an average minimum signal-to-noise ratio is met.  The spectra and calibrated by observations of 32 sky fibers, 8 reddening standard stars, and 8 spectrophotometric standard stars.  Spectra are flat-fielded and flux calibrated by the {\tt Spectro2d} pipeline.  Next, {\tt Spectro1d} identifies spectral features and classifies objects by spectral type \citep{stoughton02}.  Ninety-four percent of all SDSS quasars are identified spectroscopically by this automated calibration; the remaining quasars are identified through manual inspection.  Quasars are defined to be those extragalactic objects with broad emissions lines (full width at half maximum velocity width of $\gtrsim 1000$km s$^{-1}$, regardless of luminosity. 

Through June 2004, objects corresponding to 181 plates had been observed multiple times, with time lags between observations ranging from days to years.  As discussed in Paper I, spectra from plates observed greater than 50 days apart have not been co-added and are more suitable for use in variability studies.  There are 53 such large time-lag plate pairs containing almost 2200 quasars; 47 of these plate pairs are contained in the Third Data Release \citep[DR3;][]{abazajian05}.

\subsection{Refinement of Spectroscopic Calibration \label{calib}}

\citet[][hereafter VB04]{vandenberk04} and Paper I demonstrated that additional spectrophotometric calibration of SDSS spectra is necessary for variability studies.  We summarize here the calibration methods used in Paper I; see that work for a complete discussion. The {\tt Spectro1d} pipeline calculates three values of signal-to-noise ratio for each spectrum by calculating the median S/N ratio per pixel in the portions of the spectrum corresponding to the SDSS $g, r$ and $i$ filter transmission curves.  Hereafter, when referring to the two halves of a plate pair, we use the phrase "high-S/N epoch" to refer to the plate with the higher median $r$-band signal-to-noise ratio.  The plate with the lower median $r$-band signal-to-noise ratio will be called the "low-S/N epoch." It is worth emphasizing that this is a plate-wide designation; although most objects follow the plate-wide trend, this does not speak to the relative S/N values for any given individual object, nor does it correspond to an object's relative line or continuum flux at a given epoch.  The stars on a plate are used to resolve calibration differences between the high- and low-S/N epochs, under the assumption that the majority of stars are non-variable (precautions are taken to remove the obviously variable stars from re-calibration).  For each plate pair, we create a re-calibration spectrum, equal to the ratio of the median stellar high-S/N epoch flux to the median stellar low-S/N flux, as a function of wavelength.  This re-calibration spectrum is fitted with a 5th-order polynomial to preserve real wavelength dependences, but remove pixel-to-pixel noise (see Figure 5 of Paper I), leaving a smooth, relatively featureless curve as a function of wavelength. All low-S/N epoch spectra are then scaled by this "correction" spectrum.

\subsection{Variable Quasar Sample\label{quasars}}

In this study, we make use of the sample of variable quasars created in Paper I. Many quasars at low redshift appear as extended objects; due to the fiber nature of the spectrograph, these low-redshift objects are problematic for accurate relative spectrophotometry between epochs.  To avoid such problems with extended objects, only quasars with $z > 0.5$ were used in Paper I.  With respect to the assumed non-variable stellar population, 315 quasars were determined to have varied significantly between epochs.  These variable quasars have larger rest-frame time lags than the average for all SDSS quasars with multi-epoch spectroscopy, but are otherwise indistinguishable from the main sample. 
%Their redshifts and absolute magnitudes are similar to those of quasars which have not varied.  
This sample was first used to study the detailed dependence of variability on wavelength; for more information, see Paper I.

\section{Fitting the C~\sc{iv} Line\label{linefitting}}

\subsection{Region of Interest\label{roi}}

Using moment analysis techniques, we fit the C~\textsc{iv} line for all objects where the entire line has been observed by the SDSS spectrograph.  Both epochs are fit individually.  Thus, for each object, we obtain flux, position and profile information for the C~\textsc{iv} line at two epochs.

The line-fitting techniques used are similar to those used by \citet{vandenberk01} to fit composite spectra created from the SDSS quasar survey, with a few modifications.  \citet{vandenberk01} fit the composite spectrum C~\textsc{iv} line over the rest-frame wavelength range 1494\AA---1620\AA.  However, the C~\textsc{iv} line is typically flanked on the red side by He~\textsc{ii}$\lambda$1640 and emission from other species, such as Fe~\textsc{ii}, O~\textsc{iii}] and Al~\textsc{ii}, as well as some unidentified flux above the continuum redward of 1600\AA\ \citep[e.g.,][]{
wilkes84, boyle90,laor94,vandenberk01}.  However, over intervals centered on roughly 1480\AA\ and 1690\AA, quasar spectra are relatively free of emission, making these logical intervals to use in fitting the underlying continuum (see \S,\ref{continuum}).  

Spectra are not de-redshifted for fitting; the region of interest for each quasar is determined by scaling the [1472\AA, 1700\AA] interval by $1+z$ for that quasar.
Only those spectra containing this region in its entirety are used; quasars with redshifts between 1.65 and 4.35 are available for study.  Table \ref{civqsotable} lists these 105 objects, as well as their dates of observation (MJD), redshifts ($z$), rest-frame time lags ($\Delta{\tau}$), absolute magnitudes ($M_{r}$) and both epochs' S/N ratio.  (Absolute magnitudes are calculated assuming a power law spectral energy distribution $f_\lambda\propto \lambda^{\alpha_{\lambda}}$, with a slope of $\alpha_\lambda=-1.5$. )

% Estimated K-corrections using the composite spectrum from \citet{vandenberk01} are consistent with the simple power %law assumption, and the differences are usually no greater than 0.1 magnitude at any redshift.)

This leaves 105 objects with redshifts ranging from 1.65 to 4.00.  The fitting procedures are explained in sections \S\,\ref{continuum}---\S\,\ref{errors}.  Results of the fits are in Tables \ref{civfluxtable} and \ref{civprofiletable}.

\subsection{Continuum Fitting and Total Line Flux\label{continuum}}

Accurately fitting the underlying continuum near a line is critical.  Too low or high a continuum fit will lead to an overestimate or underestimate of the line flux.  An incorrectly fit local continuum slope could introduce an apparent asymmetry not inherent to the line itself.  Despite its importance, there is no widely accepted method of continuum fitting.  The fits to quasar continua in \citet{wills93} employed either a power-law or a low-order polynomial.  \citet{vandenberk01} fit the local continuum with a straight line.  Over such a small wavelength range ($\sim150$\AA), it appears that any reasonable function will work, provided care is taken.

For simplicity, we employ a straight-line local continuum fit.  To ensure that the fits were well behaved, numerous visual checks were done.  We varied the size of the region of interest and the size of the region used to fit the continuum.  We use here those values which appeared to give the most accurate and stable fits to the continuum and the line center.

Before fitting the continuum, the region of interest in each spectrum is manually inspected for poor night sky subtraction or absorption lines.  In the case of poor subtraction of a night sky line (most commonly O~\textsc{i}$\lambda{5577}$), we interpolate over the affected region, using the pixels within 25\AA\ on either side.  Night sky lines were removed from 8 spectra.  There are no cases where poor night sky subtraction occurs near the peak of a line.  Absorption lines are only removed in the cases where they affect the fit to the continuum; they are not removed if they lie on top of the C~\textsc{iv} emission line itself.  Only 4 absorption lines are removed. 

To fit the continuum, a single straight line is fit to pixels at either end of the region of interest.  This corresponds roughly to fitting the continuum with those pixels with rest-frame wavelengths between 1472\AA\ and 1487\AA\ or between 1685\AA\ and 1700\AA.  This linear continuum fit is then subtracted from every pixel in the region of interest to isolate the line flux:

\begin{equation}
F_{line}(\lambda)=F_{total}(\lambda)-F_{cont.}(\lambda)
\end{equation}

To avoid inclusion of emission flux from other sources, measurements are made over the range corresponding to a 100\AA\ interval centered on 1546\AA\ in the quasar's rest frame.  (Though 1549\AA\ is the laboratory wavelength of C~\textsc{iv}, the line is blueshifted in quasars such that the mean rest wavelength position is actually 1546\AA\ \citep{richards02a,vandenberk01}.)
The total flux for the line is simply the integral of the continuum-subtracted line flux density ($F_{\lambda} = F_{line}(\lambda)$) over this measurement interval:

\begin{equation}
f = \int_{C~\textsc{iv}}F_{\lambda}d\lambda,
\end{equation}
where C~\textsc{iv} indicates the interval [1496\AA,1596\AA], as described above.  
In \S\,\ref{linecenter} and \S\,\ref{width}, we calculate the first three moments of the C~\textsc{iv} line, using this continuum-free region.

\subsection{$1^{st}$ Moment: Line Center\label{linecenter}}

In a perfectly symmetric line, the meaning of the line "center" is easily understood; the mean, median and mode of the line's flux distribution all fall at the same wavelength.  It is thought, however, that the C~\textsc{iv} line is typically asymmetric \citep{wills93}.  Therefore, one must choose which statistic to use.  In the spectra used here, some of which have low signal-to-noise ratios, the median is a far more reliable measurement that the mean, which is easily affected by noisy pixels in the line wings.  Thus, for robustness and simplicity, we calculate the line center using the median.  The median is simply the midpoint of the line flux, the wavelength which evenly divides the continuum-subtracted flux in the [1496\AA,1596\AA] interval.  

We also use this measurement of the line center to calculate the local continuum; the continuum flux density is determined by evaluating the straight-line continuum fit (see \S\,\ref{continuum}) at the median-determined line center.

\subsection{$2^{nd}$ and $3^{rd}$ Moments: Line Width and Skewness\label{width}}

We use the second moment about the median wavelength as a measurement of line width:

\begin{equation}
\sigma^{2}=\frac{\int_{C~\textsc{iv}}(\lambda-\lambda_{median})^{2}F_{\lambda}d\lambda}{\int_{C~\textsc{iv}}F_{\lambda}d\lambda}.
\end{equation}

Visual inspection reveals that the Pearson Skewness,

\begin{equation}
Pskew=\frac{3\times(mean-median)}{\sigma},
\end{equation}
is a more stable statistic between epochs than the third moment of the flux.
% ($\gamma = \frac{\int_{ROI}(\lambda-\lambda_{median})^3F_{\lambda}d\lambda}{\sigma^{3}}$).  
Thus, as did \citet{vandenberk01}, we use the Pearson skewness to measure line asymmetry.  To keep the mean from being unduly affected by the noisy tails of the C~\textsc{iv} line, we measure the mean of the line over the interval [1516\AA,1576\AA], using $\lambda_{mean} = \frac{\int \lambda F_{\lambda}d\lambda}{\int F_{\lambda}d\lambda}$.  This is tantamount to \citet{vandenberk01} using only the "top $50\%$" of the line.  The single-object spectra used here are too noisy to allow for a reliable determination of the "top $50\%$", so this 60\AA\ interval, chosen because it closely approximates the "top 50\%", is used instead.

\begin{figure}
\includegraphics[scale=1.0]{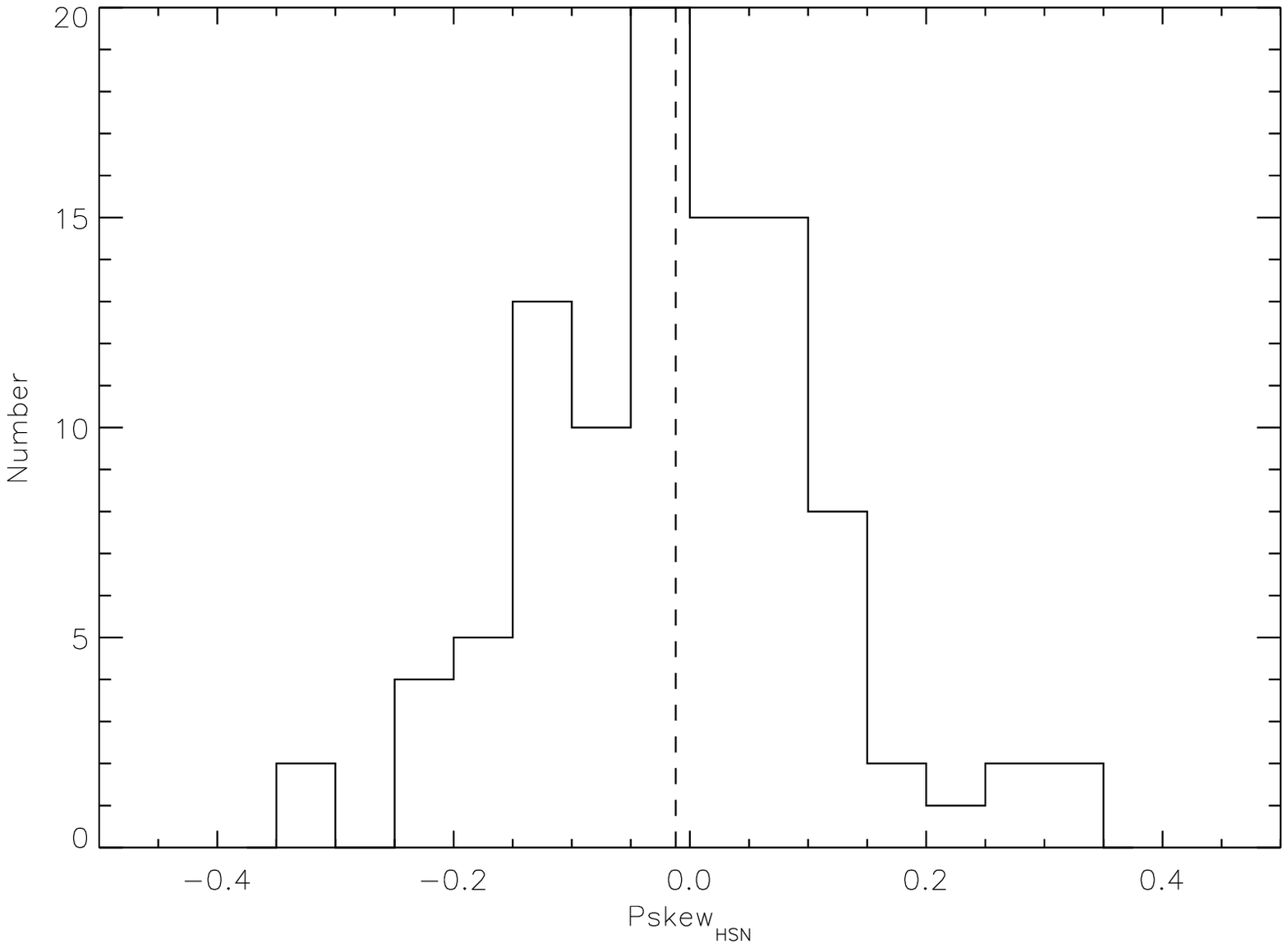}
\caption{Distribution of Pearson skewness for quasars at high-S/N epoch.  The median skewness is $0.012 \pm 0.013$.
\label{pskew1hist}}
\end{figure}

%By substituting the relationship in equation 6, equation 8 may be rewritten as $Pskew=\frac{3\times(median-mode)}{2\sigma}$.  In this form, it is easy to see that a line with negative skewness is one where $\lambda_{median} < \lambda_{mode}$.  Therefore, lines of negative skewness  are those with the majority of the flux blueward of the line peak.  
Fig. \ref{pskew1hist} is a histogram of the high-S/N epoch Pearson skewness values for the objects in our sample.  The median value of $-0.012 \pm 0.013$ is consistent with no skewness of the C~\textsc{iv} line.  \citet{vandenberk01} measured a C~\textsc{iv} Pearson skewness of -0.04 for the SDSS composite quasar spectrum.  The lower value measured here is likely due to the difference in fitting the continuum.  Using the region around 1690\AA\ to fit the red side of the continuum leads to a bluer continuum and, therefore, the inclusion of more flux from the red side of the C~\textsc{iv} line than either \citet{vandenberk01} or \citet{wills93}.

\subsection{Line Flux Requirement\label{highflux}}

Accurate measurement of line width, skewness and C~\textsc{iv}-Mg~\textsc{ii} line shift requires accurate and precise determination of the line center.  When searching for changes in these quantities with time, it is essential to have reliable measurements at both epochs.  Visual inspection indicates that the fitting procedure returns reasonable parameter values for those objects with flux $f_{line,HSN} > 200 \times 10^{-17}$ erg/s/cm$^{2}$.  We thus remove those 10 objects below this limit from future study.  For reference, we do include the results of the fitting code for these objects in Tables \ref{civfluxtable} and \ref{civprofiletable}.

\subsection{Errors in Fitted Quantities\label{errors}}

%In principle, the errors in the various moments may be calculated in a straightforward manner.  The error in the centroid, for example, may be determined by %integrating the error in the flux over the region of interest:
%should represent the 

%\begin{equation}
%\sigma_{\lambda}^{2}=\frac{\int_{ROI}\sigma_{F_{\lambda}}^{2}(\lambda-\lambda_{centroid})^{2}d\lambda}{\int_{ROI}F_{\lambda}d\lambda}
%\end{equation}

%In practice, however, the errors returned by this method were unreasonably large.  The errors in the line centers, for example, were estimated to be $\sim$100\AA ($\sigma_{z}\sim0.06-0.07$).  As the typical C~\textsc{iv} line width is $\sim$ 50-60\AA, these errors were at least an order of magnitude too large.  

%Instead of using these overestimated errors, w
We use a Monte Carlo method for determination of errors.  We add random, Gaussian noise to each pixel in the region of interest by assigning a random number drawn from a Gaussian distribution with mean equal to the measured flux in that pixel and standard deviation equal to the measured error in that pixel.  The moment analysis code is run on this altered spectrum and values for continuum flux, line flux, center, width, and skewness are calculated.  This is repeated 1000 times; the error assigned to each quantity is equal to the standard deviation of the distribution of resulting values.  

At the high-S/N epoch, the median errors in the line flux and the continuum flux density are $33 \times 10^{-17}$ erg/s/cm$^{2}$ and $0.093 \times 10^{-17}$ erg/s/cm$^{2}$/\AA, repsectively.  The median error in the line center (as determined by the median) is 1.2\AA.  This is less than half the value for any of the other methods used to measure the line center.  The medians in the line width and skewness error distributions are 2.8\AA\ and 0.05, respectively.

\subsection{Mg~\sc{ii} Line Fitting\label{mgii}}

For use in studying the effect of variability on the C~\textsc{iv}-Mg~\textsc{ii} line shift, we also fit the Mg~\textsc{ii} line for those objects where that line also falls in the SDSS spectroscopic wavelength range.  For Mg~\textsc{ii}, the region of interest for each line was defined to be those pixels which correspond to rest-frame wavelengths between 2684\AA\ and 2913\AA, the same range used in \citet{vandenberk01}.  Requiring SDSS coverage of the entire region of interest means that the Mg~\textsc{ii} line is fit to those 77 objects in the C~\textsc{iv} sample with $z < 2.12$.  The Mg~\textsc{ii} line is fit at both the high- and low-S/N epochs, using the same algorithm as the C~\textsc{iv} lines.  It should be noted that there is significant Fe~\textsc{ii} emission on either side of the Mg~\textsc{ii} line \citep[e.g.,][]{vestergaard01,sigut03,baldwin04}.  No attempt has been made in this work to remove this Fe~\textsc{ii} flux, as only the robust Mg~\textsc{ii} line median measurement is used.  In any detailed study of the Mg~\textsc{ii} line profile, however, this Fe~\textsc{ii} emission should be removed.  See \S\,\ref{lineshifts} for a discussion of variability of the blueshift of the C~\textsc{iv} line relative to Mg~\textsc{ii}.

\section{Variability of the C~\sc{iv} Line\label{civvariability}}

\subsection{Profile\label{profile}}

\begin{figure}
\includegraphics[scale=1.0]{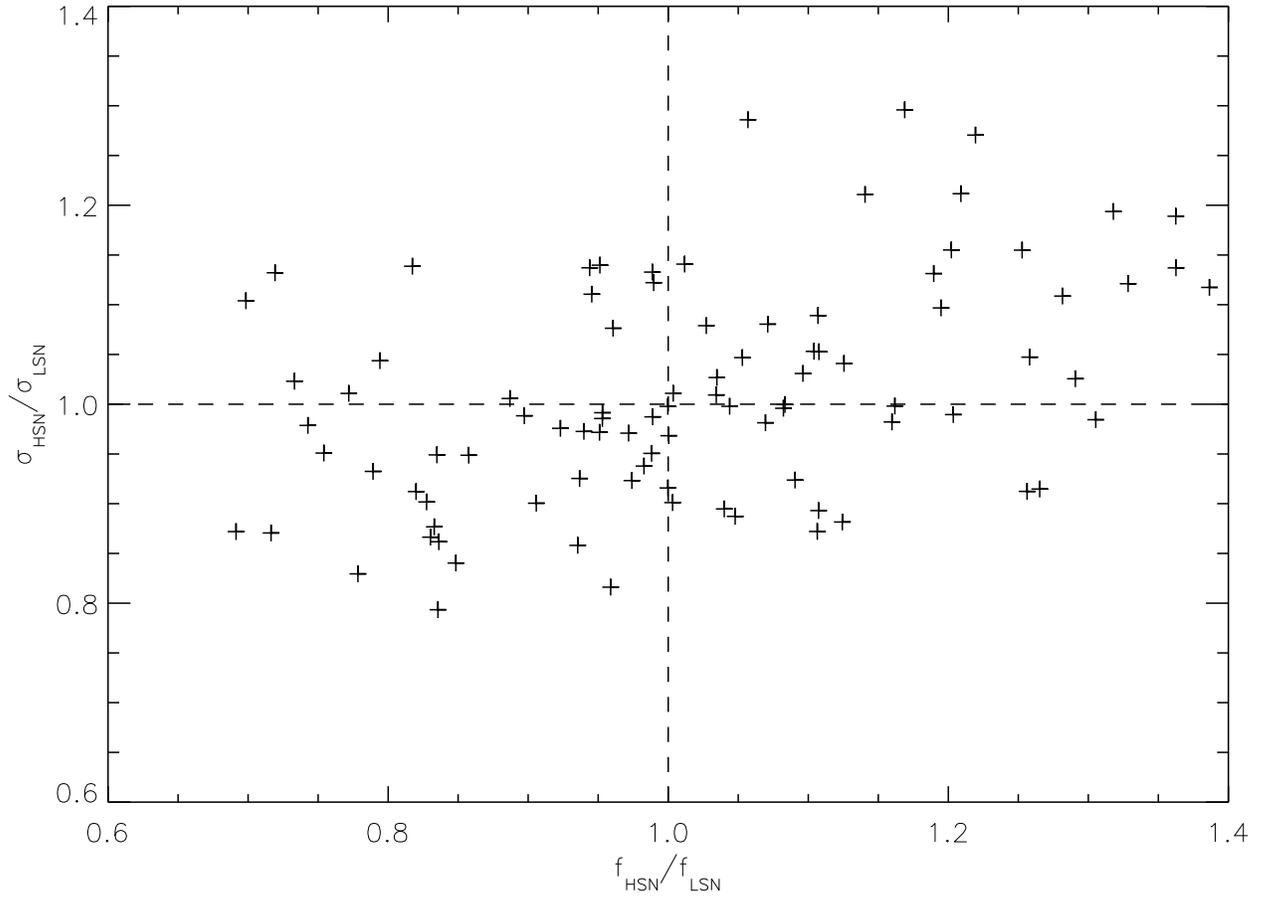}
\caption{Change in C~\textsc{iv} line width versus line flux change.  The Spearman rank correlation coefficient for this distribution is 0.42 with a significance of $2.3 \times 10^{-5}$.
\label{df.dsigma}}
\end{figure}

There appears to be a strong correlation between the change in the line flux and the change in the line width.  Fig. \ref{df.dsigma} shows the epoch-to-epoch flux ratio ($\frac{f_{HSN}}{f_{LSN}}$) versus the ratio of line widths ($\frac{\sigma_{HSN}}{\sigma_{LSN}}$).  A Spearman rank correlation test yields a correlation coefficient of 0.42 with a significance of $2.3\times10^{-5}$.  The Kendall correlation coefficient is 0.29 with a significance of $2.5\times10^{-5}$.  These tests indicate there is a very low probability that no correlation exists.  This strongly suggests that, for an individual object, the C~\textsc{iv} line width increases with the line flux.  This is opposite the sense of the single-epoch anticorrelation between C~\textsc{iv} equivalent width and FWHM seen by \citet{wills92}.  It appears that there are two separate relations between line strength and line width: a global relation, like that seen by \citet{wills92} and \citet{francis92}, which suggests that, from object to object, line strength is anti-correlated with line width; and an intrinsic relation seen in Fig. \ref{df.dsigma} which suggests that line width and line strength are correlated for an individual object.

\begin{figure}
\includegraphics[scale=1.0]{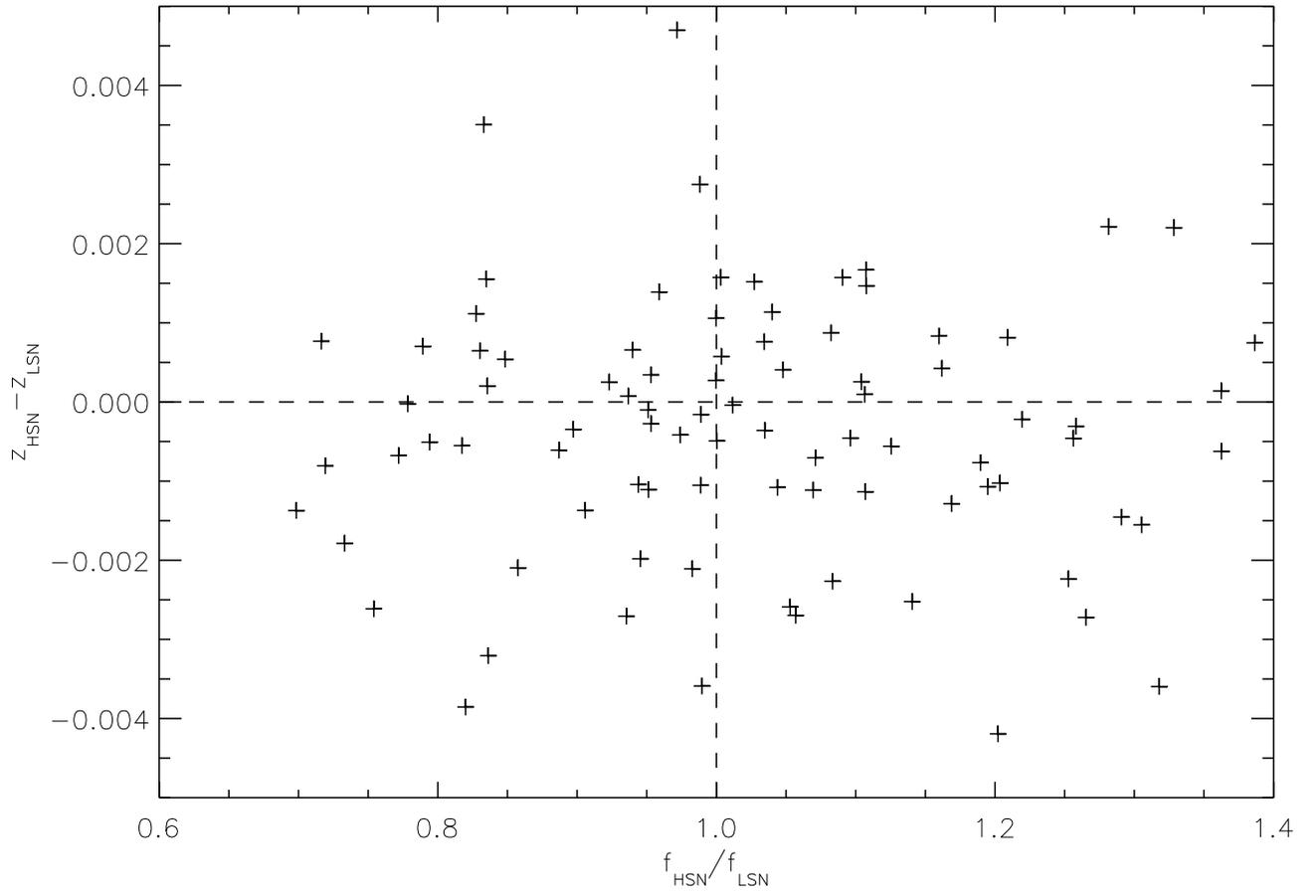}
\caption{Change in C~\textsc{iv} line center versus line flux change.  The Spearman rank correlation significance for this distribution is 0.86. 
\label{df.dz}}
\end{figure}

\begin{figure}
\includegraphics[scale=1.0]{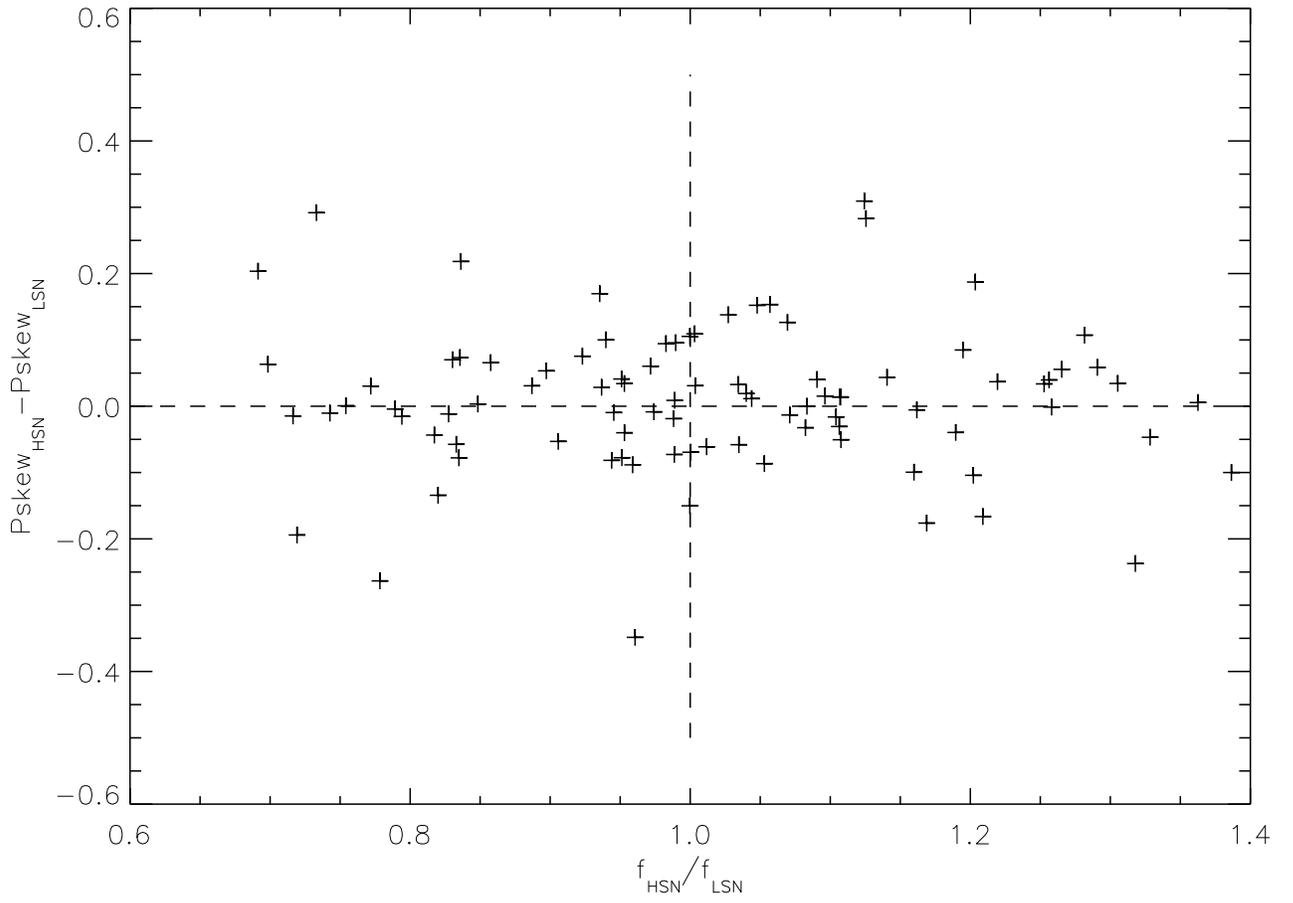}
\caption{Change in C~\textsc{iv} line Pearson skewness versus line flux change.  The Spearman rank correlation significance for this distribution is 0.34.
\label{df.dpskew}}
\end{figure}

There is no obvious correlation between C~\textsc{iv} line flux and either line center or skewness.  Fig. \ref{df.dz} shows the epoch-to-epoch flux ratio ($\frac{f_{HSN}}{f_{LSN}}$) versus the change in median-determined redshift ($\Delta{z} = z_{HSN} - z_{LSN}$) and Fig. \ref{df.dpskew} shows the flux ratio versus the change in Pearson skewness ($\Delta{Pskew} = Pskew_{HSN} - Pskew_{LSN}$).  No trend is apparent in either plot.  This is reinforced by the Spearman rank correlation tests.  The Spearman significances are 0.86 and 0.734 for the flux-redshift and flux-skewness distributions, respectively, indicating that there is no significant correlation with flux change for either redshift change or skewness change.

\subsection{Flux\label{flux}}
\subsubsection{Luminosity and Time Dependences\label{lumtime}}

\begin{figure}
\includegraphics[scale=1.0]{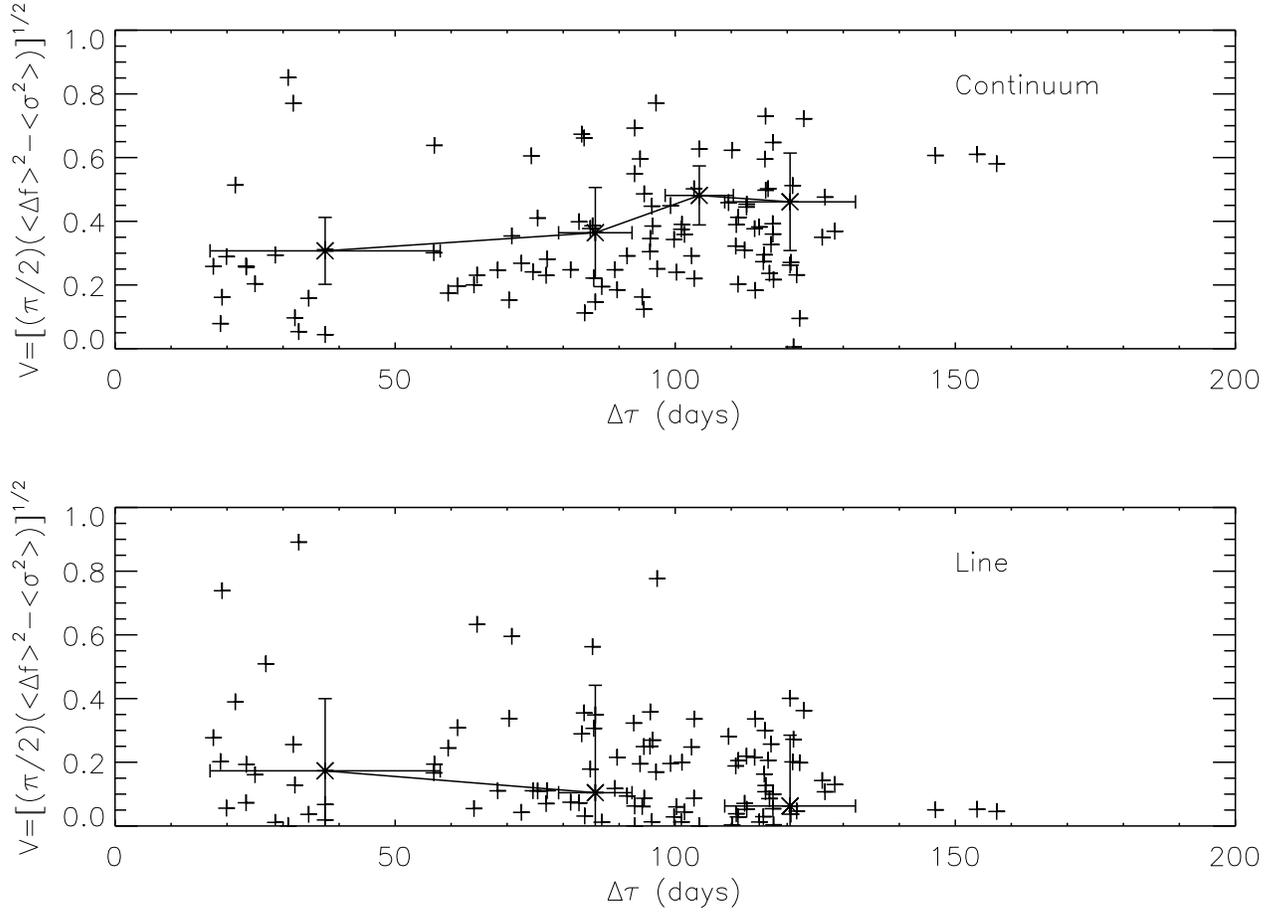}
\caption{Continuum flux (upper panel) and C~\textsc{iv} line flux (lower panel) variability (V) as a function of rest-frame time lag ($\Delta{\tau}$).  The overlaid binned points correspond to the well-known Structure Function.
\label{civbothsfs}}
\end{figure}

\begin{figure}
\includegraphics[scale=1.0]{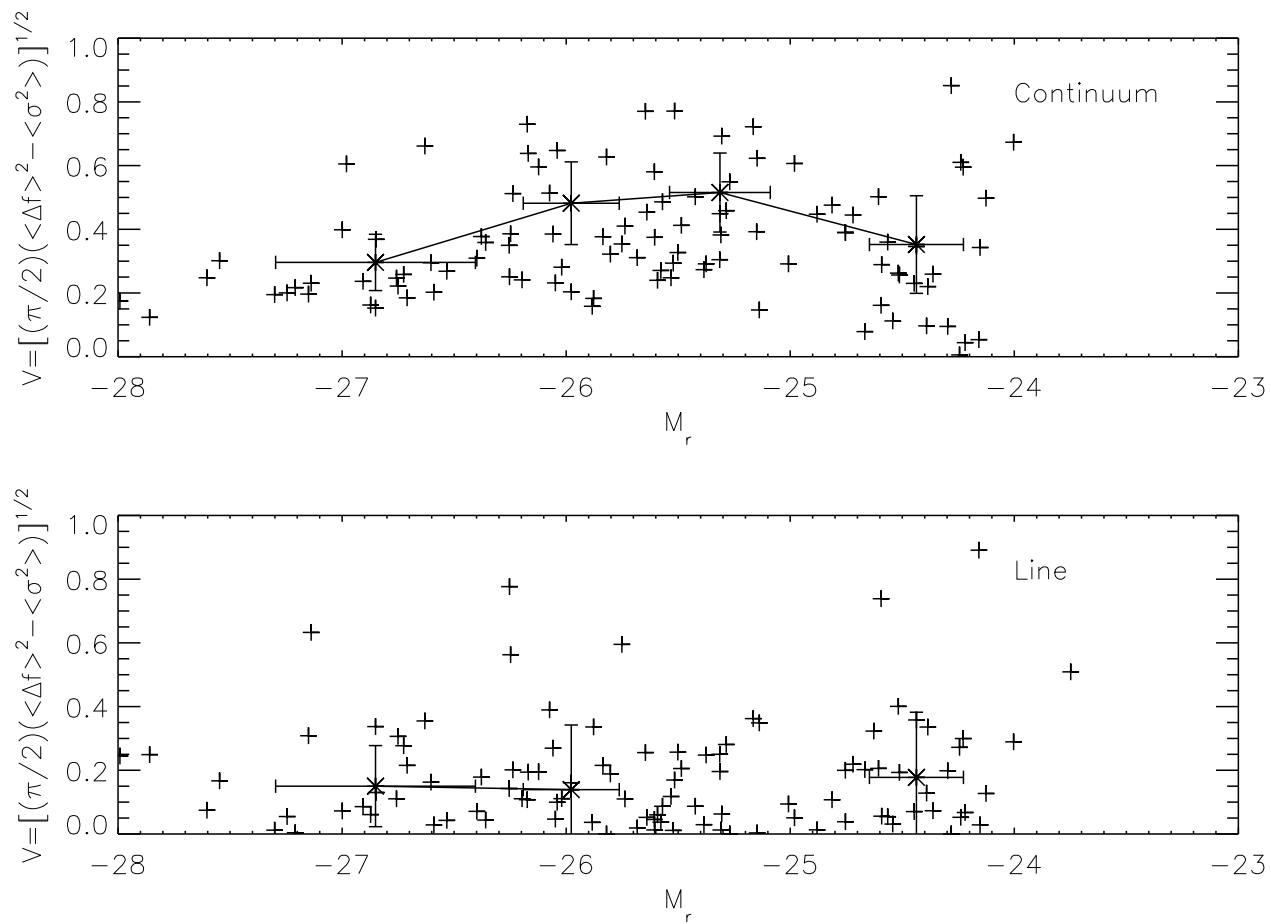}
\caption{Continuum flux (upper panel) and C~\textsc{iv} line flux (lower panel) variability (V) as a function of $r$-band absolute magnitude.  Overlaid are the binned values for V and $M_{r}$.  Large errors in the individual values of V in the lowest-luminosity bin make the binned measurements of V unreliable.
\label{civbothmr}}
\end{figure}

The upper panels of Figs. \ref{civbothsfs} and \ref{civbothmr} show the well-known dependence of continuum variability amplitude upon rest-frame time lag and absolute magnitude, respectively. To make comparisons with past photometric studies easier we measure the change in flux between epochs by computing the logarithm of the ratio of the fitted continuum fluxes at the two epochs for each object: $\Delta{f} = -2.5 log(\frac{f_{cont,HSN}}{f_{cont,LSN}})$.  The error in $\Delta{f}$, $\sigma_{\Delta{f}}$, is calculated for each object through standard error propagation: $\sigma_{\Delta{f}} = \frac{2.5}{ln 10}\sqrt{(\frac{\sigma_{f_{cont,HSN}}}{f_{cont,HSN}})^{2} + (\frac{\sigma_{f_{cont,LSN}}}{f_{cont,LSN}})^{2}}$.

We then create four equally populated bins in rest-frame time lag and absolute magnitude.  In each bin, we calculate the average continuum variability by removing the average flux change error ($\sigma_{\Delta{f}}$) from the average flux change ($\Delta{f}$): 

\begin{equation}
V = \sqrt{(\frac{\pi}{2})<\Delta{f}>^2-<\sigma_{\Delta{f}}^2>},
\end{equation}
as was done in VB04.  Because of the small number of objects in each bin ($\sim25$), we use the median for the average (instead of the mean as in VB04).  The rest-frame time-lag dependence of variability is commonly referred to as the structure function.  
The upper panel of Fig. \ref{civbothsfs} shows the continuum variability ($|\Delta{f}|$) versus rest-frame time lag for all individual objects, with the binned ensemble structure function overlaid.  Similarly, the upper panel of Fig. \ref{civbothmr} shows the binned ensemble luminosity dependence overlaid on the individual objects' continuum variability versus luminosity.

Qualitatively, the structure function in the upper panel of Fig. \ref{civbothsfs} is similar to that seen in previous studies of continuum variability \citep[see VB04;][]{hook94}.  The structure function increases with time, indicating that quasar continua are more likely to appear to have varied when the time interval between observations is long, as seen in \citet{rengstorf05}.
The well-known \citep[e.g. VB04;][]{giveon99} anti-correlation between variability amplitude and quasar luminosity is seen in the upper panel of Fig. \ref{civbothmr}; more luminous quasars tend to exhibit less continuum variability.  The bin containing the intrinsically faintest objects does not follow this trend; it is unlikely that this dip in variability amplitude is statistically significant.  With such a small number of objects per bin, this measure of the variability is easily affected by large or overestimated errors in the variability of individual objects.  To avoid this problem, VB04 required that each bin contain a minimum of 75 objects; with only 94 total objects, we do not have this luxury here.
Quantitatively, the amplitude of the variability in the upper panels of Figs. \ref{civbothsfs} and \ref{civbothmr} is larger than that of previous studies.  This can be at least partially understood as an artifact of the creation of the sample; only those quasars which had demonstrated significant variability were chosen for study in Paper I.

The well-known relationships between continuum variability amplitude and time lag and luminosity do not appear to hold for the C~\textsc{iv} line flux.  Replacing continuum flux with line flux, we again calculate the relative flux change between epochs ($\Delta{f} = -2.5 log(\frac{f_{line,HSN}}{f_{line,LSN}})$), as well as the error in the relative flux change ($\sigma_{\Delta{f}} = \frac{2.5}{ln 10}\sqrt{(\frac{\sigma_{f_{line,HSN}}}{f_{line,HSN}})^{2} + (\frac{\sigma_{f_{line,LSN}}}{f_{line,LSN}})^{2}}$).  We then calculate the line flux variability, V, in each of the four bins in L and $\Delta{\tau}$ and $M_{r}$, using equation 9.  These are plotted in the lower panels of Figs. \ref{civbothsfs} and \ref{civbothmr}.
The lower panel of Fig. \ref{civbothsfs} appears to show a {\it decrease} in line variability amplitude with rest-frame time lag.  The error bars (representing the standard deviation of the $\Delta{f}$ distribution in each bin) are quite large, though, indicating that this decrease is not statistically significant.  Regardless, the variability amplitude does not obviously demonstrate the same time lag dependence as the continuum variability amplitude.  One is cautioned not too read too much into this lack of dependence, however.  We have no knowledge of the quasars' individual light travel times from the central source to the C~\textsc{iv}-emitting portion of the BLR.  A true C~\textsc{iv} structure function would likely require some correction for the light travel time to the BLR, something impossible to obtain from only-two epoch data.  Thus. without this correction, is perhaps unsurprising that no dependence is evident.  As seen in the lower panel of Fig. \ref{civbothmr}, the line variability luminosity dependence also fails to duplicate the relation seen in the continuum variability.  

\subsubsection{The Baldwin Effect and the Internal Baldwin Effect\label{Baldwin}}

\begin{figure}
\includegraphics[scale=1.0]{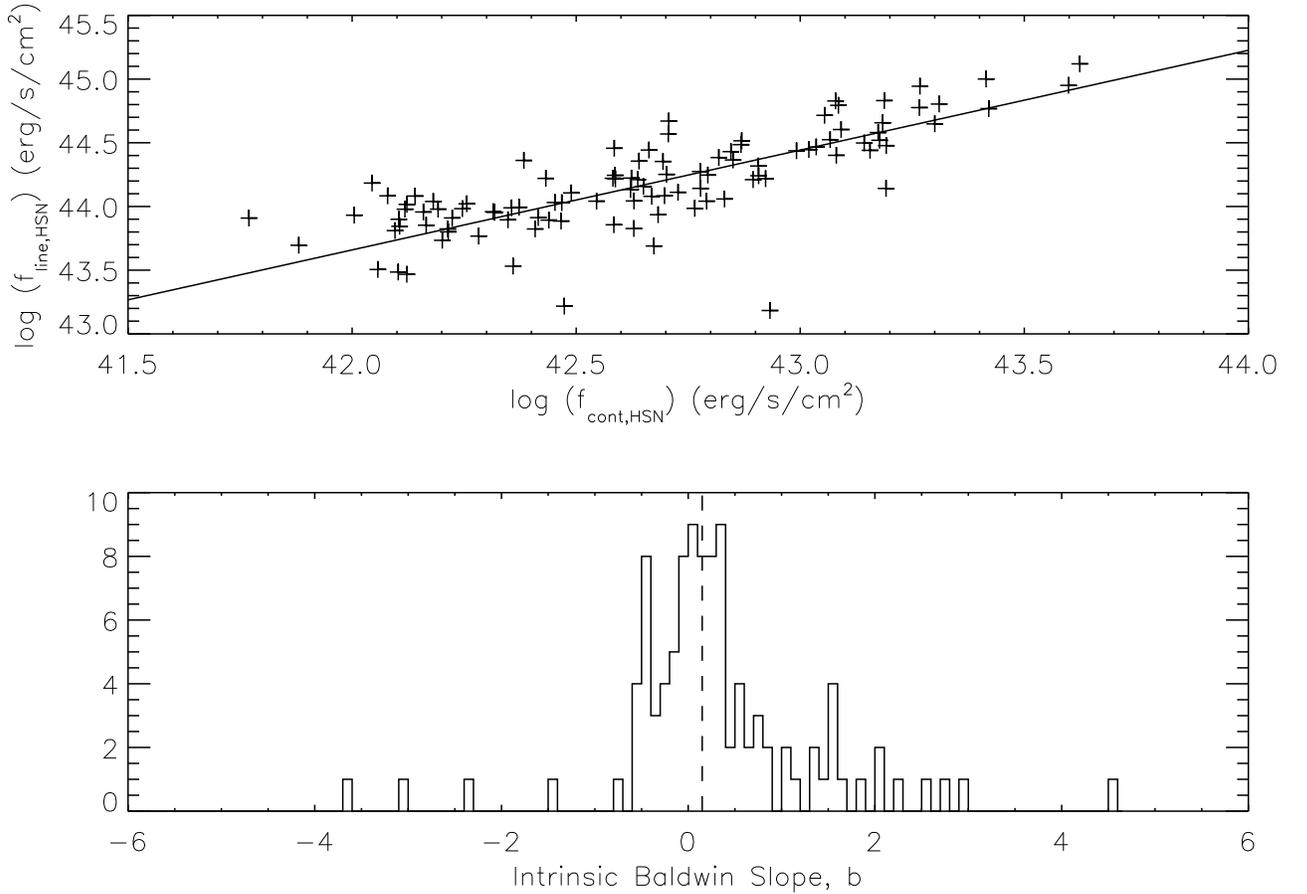}
\caption{C~\textsc{iv} Line flux versus continuum flux (upper panel) for objects at high-S/N epoch display the familiar Baldwin Effect.  (lower panel) Histogram of C~\textsc{iv} Intrinsic Baldwin Effect slopes.  The median slope (after untrustworthy outliers are removed) $b_{int}=0.15 \pm 0.06$.
\label{baldwin2panel}}
\end{figure}

Using the fit values for the continuum flux at the C~\textsc{iv} line center and the flux of the line itself determined in \S\,\ref{linefitting}, we are able to recreate the Baldwin Effect.  The upper panel of Fig. \ref{baldwin2panel} shows continuum luminosity versus line luminosity for all 105 objects at the high-S/N epoch.  Using the IDL routine {\tt POLYFITW}, the power-law slope of this relation is measured to be $b=0.78 \pm 0.03$ for the high-S/N epoch (and $b=0.82 \pm 0.03$ for the low-S/N epoch), corresponding to $\beta=-0.22$ (and $\beta=-0.18$) for the equivalent width formulation of equation 1.  These values are in rough agreement with the $\beta=-0.17 \pm 0.04$ value measured by \citet{kinney90}.

Combining the data from both epochs, we calculate an Intrinsic Baldwin Effect (IBE) slope for every object in the sample:

\begin{equation}
b_{int} = \frac{log(L_{line,HSN}/L_{line,LSN})}{log(L_{cont,HSN}/L_{cont,LSN})}
\end{equation}
The calculated IBE slope for any one object should not be taken as a definitive measurement of the IBE slope for that object, since it is determined with data from only two epochs, but the distribution of IBE slopes should be meaningful.  This distribution is quite wide, as is seen in the lower panel of Fig. \ref{baldwin2panel}.  The median IBE slope of the entire sample is $b_{int}=0.15 \pm 0.49$.  The error in the mean is large because of the few objects with large values for the IBE slope.  These large slopes are not likely to be trustworthy (a very small change in continuum luminosity between epochs can lead to a very large, but essentially meaningless, value for the Intrinsic Baldwin Effect slope).  Excluding those 12 objects with an absolute value for the IBE slope greater than 2, we find $b_{int}=0.15 \pm 0.06$.  This is shallower than the $b_{int} \approx 0.35$ value found by \citet{kinney90}.  The small number of objects (6) in the \citet{kinney90} sample precludes us from drawing strong conclusions about this difference.  However, if this difference is real, it is an indication that the C~\textsc{iv} lines of high-luminosity quasars are less responsive to continuum variations than those of low-luminosity quasars, in agreement with the results of \citet{kaspi03}.

%There is no clear dependence of IBE slope upon luminosity for objects within this sample.  Figure \ref{ibe.mr} shows Intrinsic Baldwin slope versus %absolute $r$-band magnitude for all individual objects in the sample.  Overlaid are the median IBE slope for four equally populated bins in $M_{r}$ %(error bars displayed represent the standard deviation in IBE slope for objects in each bin).  No trend is apparent; within errors, the values from all %bins agree.

\subsection{Line Shifts\label{lineshifts}}

\begin{figure}
\includegraphics[scale=1.0]{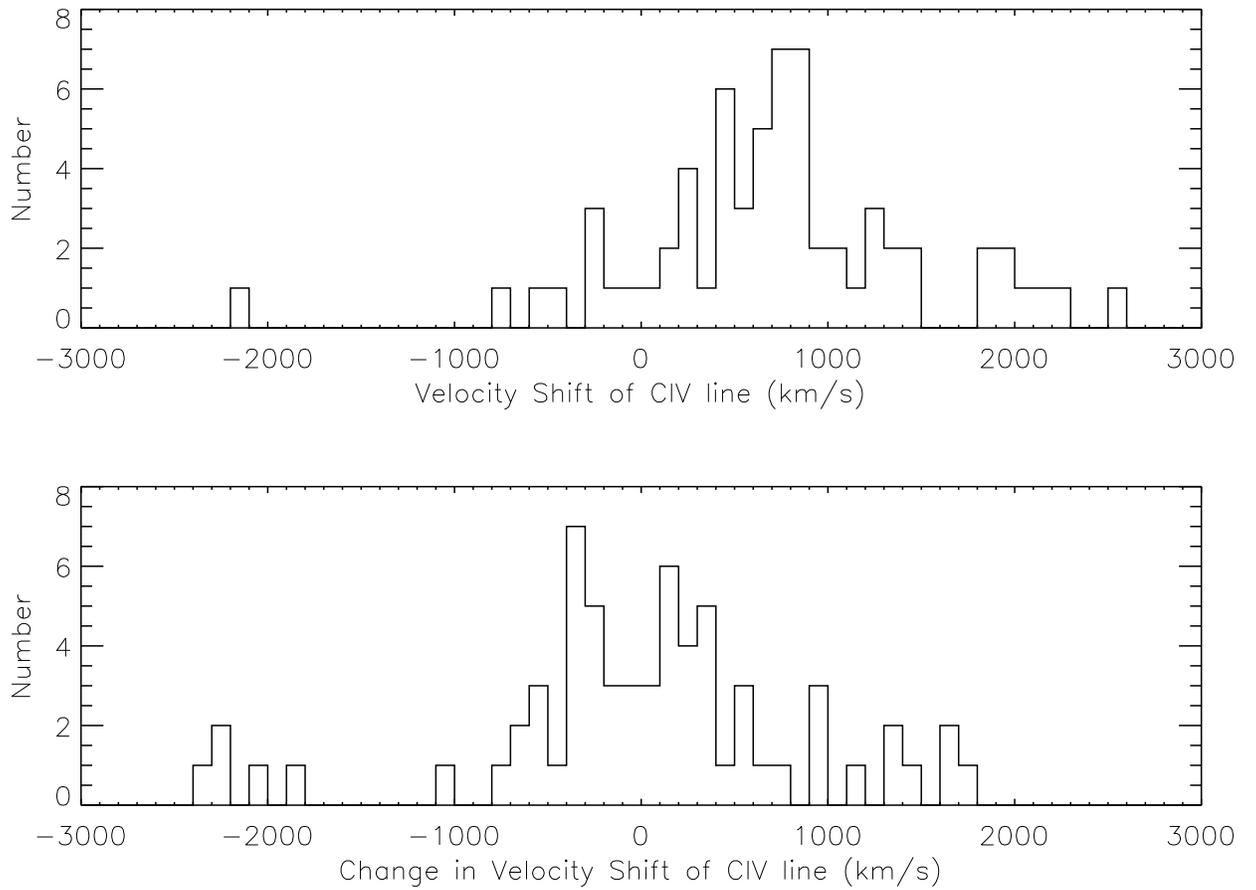}
\caption{Upper panel: High-S/N epoch velocity shift between C~\textsc{iv} and Mg~\textsc{ii} line centers.  Positive velocities indicate a C~\textsc{iv} blueshift.  Lower panel: Difference in C~\textsc{iv}-Mg~\textsc{ii} line shift between the high- and low-S/N epochs.
\label{lineshift2panel}}
\end{figure}

We are able to reproduce the single-epoch C~\textsc{iv} line shift with respect to Mg~\textsc{ii}, seen recently by \citet{richards02a}.  The upper panel of Fig. \ref{lineshift2panel} shows a histogram of these line shifts at the high-S/N epoch.  In this representation, a positive velocity ($v_{HSN} > 0$ km/s) indicates a blueshift of C~\textsc{iv} with respect to Mg~\textsc{ii}.  The median line shift for our 77 objects is 722 km/s, with a standard deviation of 1750 km/s.  With over 700 objects, \citet{richards02a} found a median blueshift of 824 km/s and a dispersion of 511 km/s.  

The lower panel of Fig. \ref{lineshift2panel} shows the differences in line shift between the two epochs ($\Delta{v} = v_{HSN}-v_{LSN}$).  The median line shift difference is near 0; $<\Delta{v}>$ = 36 km/s, but the width of the $\Delta{v}$ distribution (1680 km/s) is almost as wide as the distribution of single-epoch line shifts.

\begin{figure}
\includegraphics[scale=1.0]{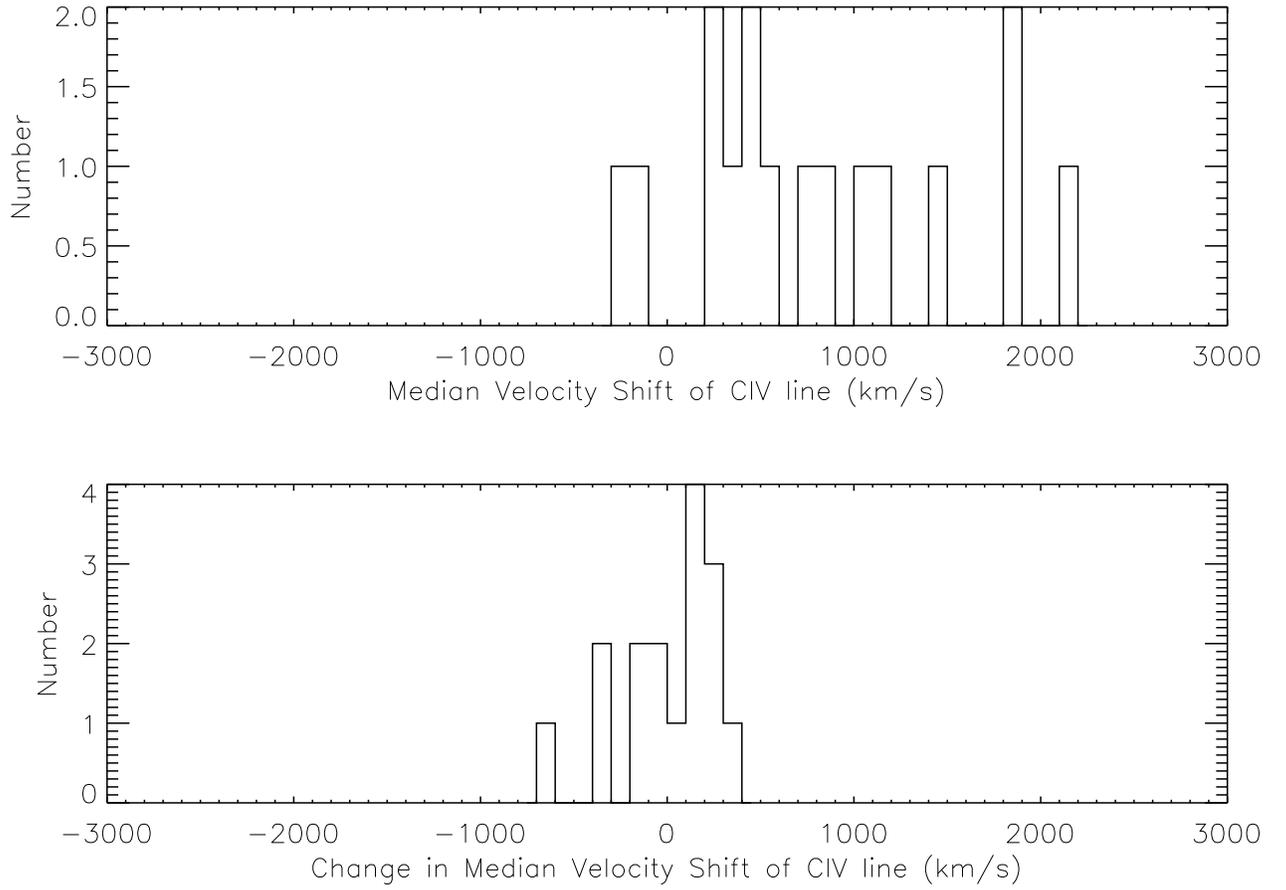}
\caption{Upper panel: Same as Fig. \ref{lineshift2panel}, for objects with $f_{line,HSN} > 800 \times 10^{-17}$ erg/s/cm$^{2}$.
\label{lineshift2panelhighflux}}
\end{figure}

Fig. \ref{lineshift2panelhighflux} is is similar to Fig. \ref{lineshift2panel}, but only includes those objects with the highest C~\textsc{iv} line flux ($f_{line,HSN} > 800 \times 10^{-17}$ erg/s/cm$^{2}$).  
This high line flux sample shows a similar median line shift (719 km/s), and a smaller standard deviation (710 km/s) than the sample as a whole.  
%That this is a smaller line shift than was found for the entire sample is not a surprise; \citet{richards02a} found that the C~\textsc{iv}---Mg~\textsc{ii} line shift decreases with increasing line luminosity.

The lower panel of Fig. \ref{lineshift2panelhighflux} shows the distribution of differences in line shift between epochs for the high line flux sample.  This distribution has a median of 120 km/s and dispersion of 280 km/s, much narrower than its counterpart for the entire sample.   
As the distribution of line shift differences for these objects is essentially centered at 0 km/s and has a much narrower width than the single-epoch distribution of line shifts (300 km/s vs. 940 km/s), it is consistent with no difference in C~\textsc{iv}-Mg~\textsc{ii} line shift between epochs.  As will be discussed in \S\,\ref{discussion}, this is further indication that C~\textsc{iv} line shifts with respect to Mg~\textsc{ii} might be useful as an orientation measure for radio-quiet quasars, if a high enough line flux can be obtained.

\section{Individual Objects\label{individual}}

\subsection{High Line-Flux Objects\label{highfluxexamples}}

We have selected some of the highest C~\textsc{iv} line flux objects in our sample to illustrate the relationship between line flux change and line width change.   SDSS J115154.83$-$005904.6 was selected based on odd change in C~\textsc{iv} line profile (see \ref{SDSSJ115154.83-005904.6}).  In all Figures, the dark curve represents the spectrum from the high-S/N epoch, the light curve the low-S/N epoch.

\begin{figure}
\includegraphics[scale=1.0]{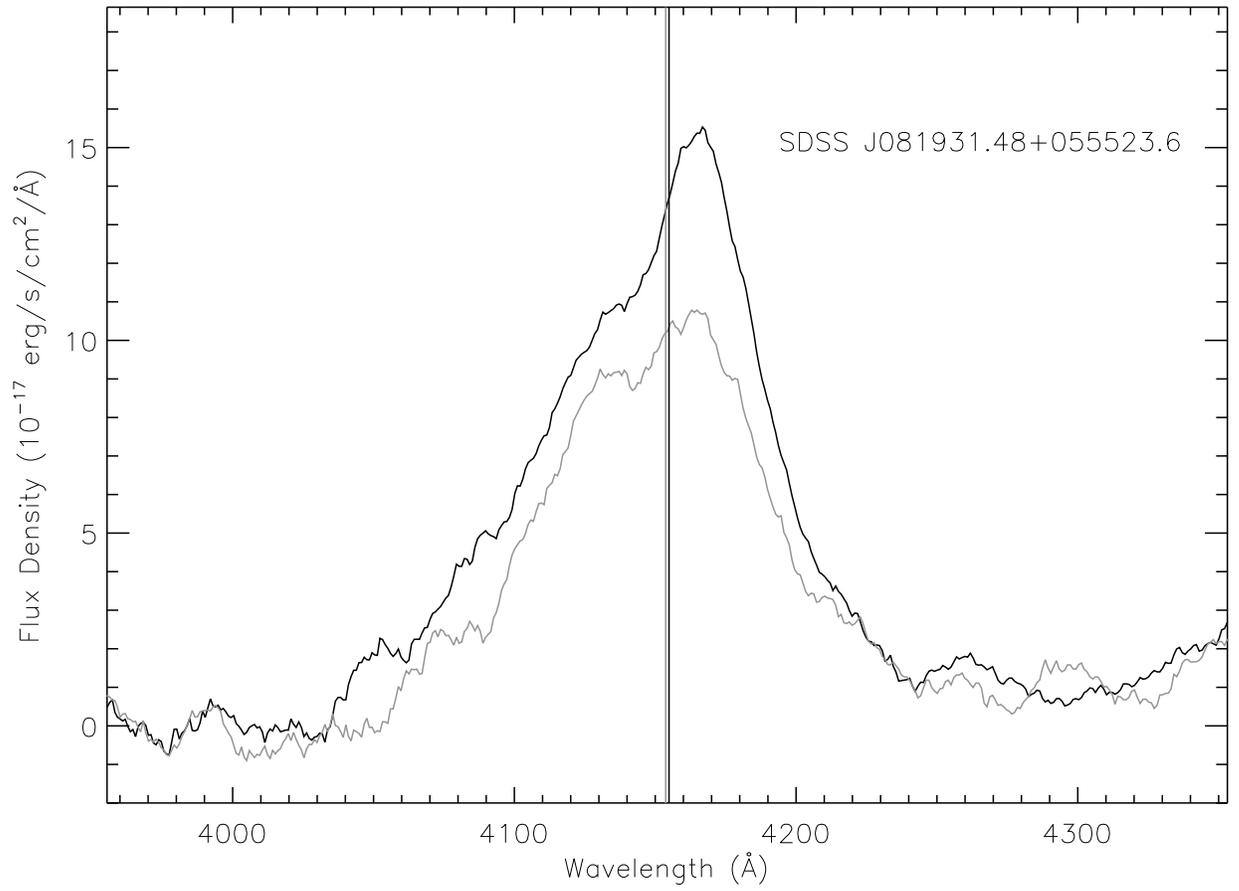}
\caption{C~\textsc{iv} line for quasar SDSS J081931.48+055523.6 at the high-S/N epoch (dark spectrum) and the low-S/N epoch (light).  Vertical lines indicate median fits to line centers.
\label{SDSSJ081931.48+055523.6fig}}
\end{figure}

SDSS J081931.48+055523.6 (Fig. \ref{SDSSJ081931.48+055523.6fig}) has line flux ratio $\frac{f_{HSN}}{f_{LSN}} = 1.39$ and width ratio $\frac{\sigma_{HSN}}{\sigma_{LSN}} = 1.12$.

%and skewness difference $Pskew_{HSN} - Pskew_{LSN} = -0.49$.  The changes in redshift and skewness are due largely to an increase in flux blueward of the peak.  It appears, in fact, that the peak itself shifts little between epochs, if at all.  As the line blue flux increases relative to the red, the median shifts and the line becomes more negatively skewed.  

\begin{figure}
\includegraphics[scale=1.0]{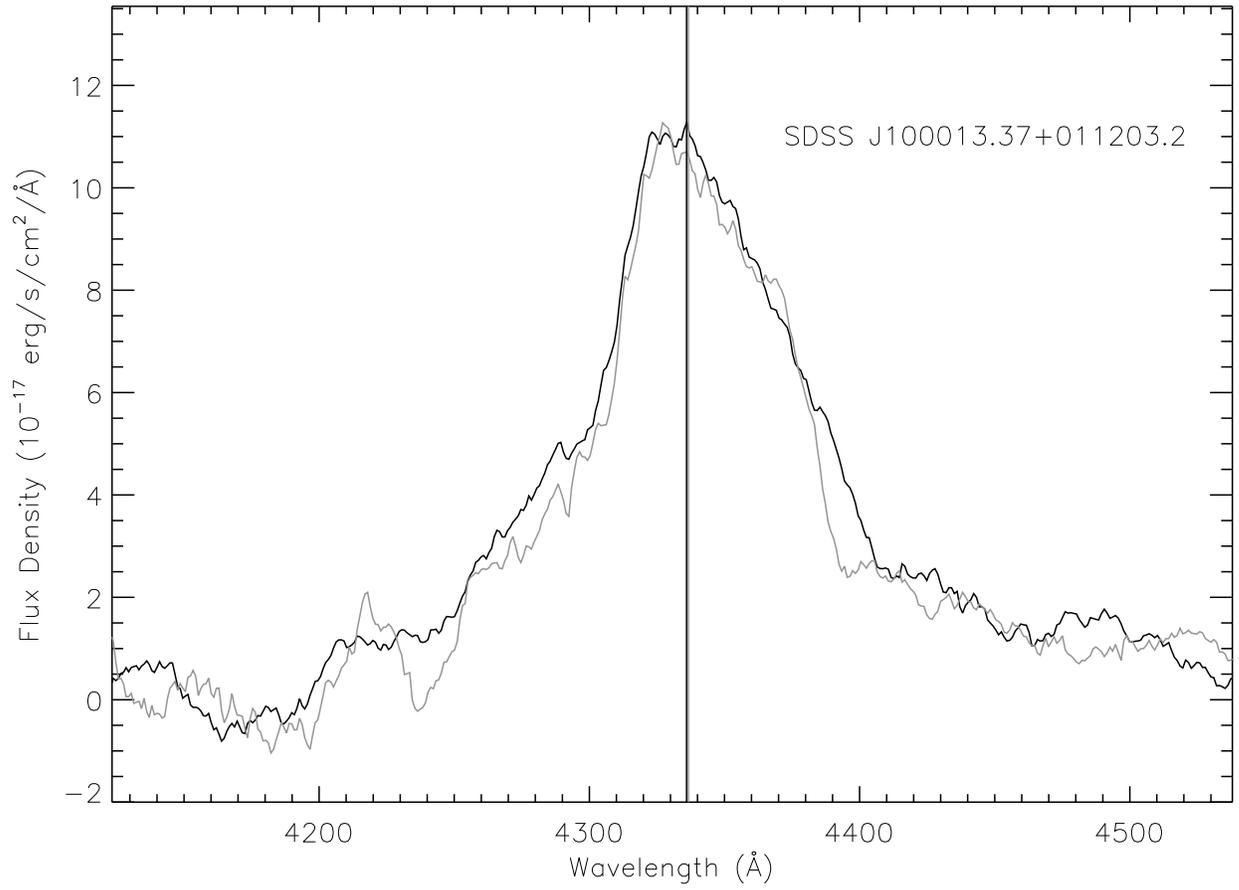}
\caption{C~\textsc{iv} line for quasar SDSS J100013.37+011203.2.  
\label{SDSSJ100013.37+011203.2fig}}
\end{figure}

SDSS J100013.37+011203.2 (Fig. \ref{SDSSJ100013.37+011203.2fig}) has line flux ratio $\frac{f_{HSN}}{f_{LSN}} = 1.09$ and width ratio $\frac{\sigma_{HSN}}{\sigma_{LSN}} = 1.03$.
%As the peak flux is not appreciably larger in the bright phase, the increase in line flux seems to be due entirely to the increased line width.

%SDSS J082328.61+061146.0 (Fig. \ref{SDSSJ082328.61+061146.0fig}) has line flux ratio $\frac{f_{HSN}}{f_{LSN}} = 1.15$ and redshift difference $z_{HSN} - z_{LSN} = -0.004$.  

%The median at the high-S/N epoch (dark curve) is shifted blueward from the low-S/N median.  This shift seems to be largely, if not entirely due to increased flux on the side of the peak, as the two phases have nearly identical spectra redward of the peak.

\begin{figure}
\includegraphics[scale=1.0]{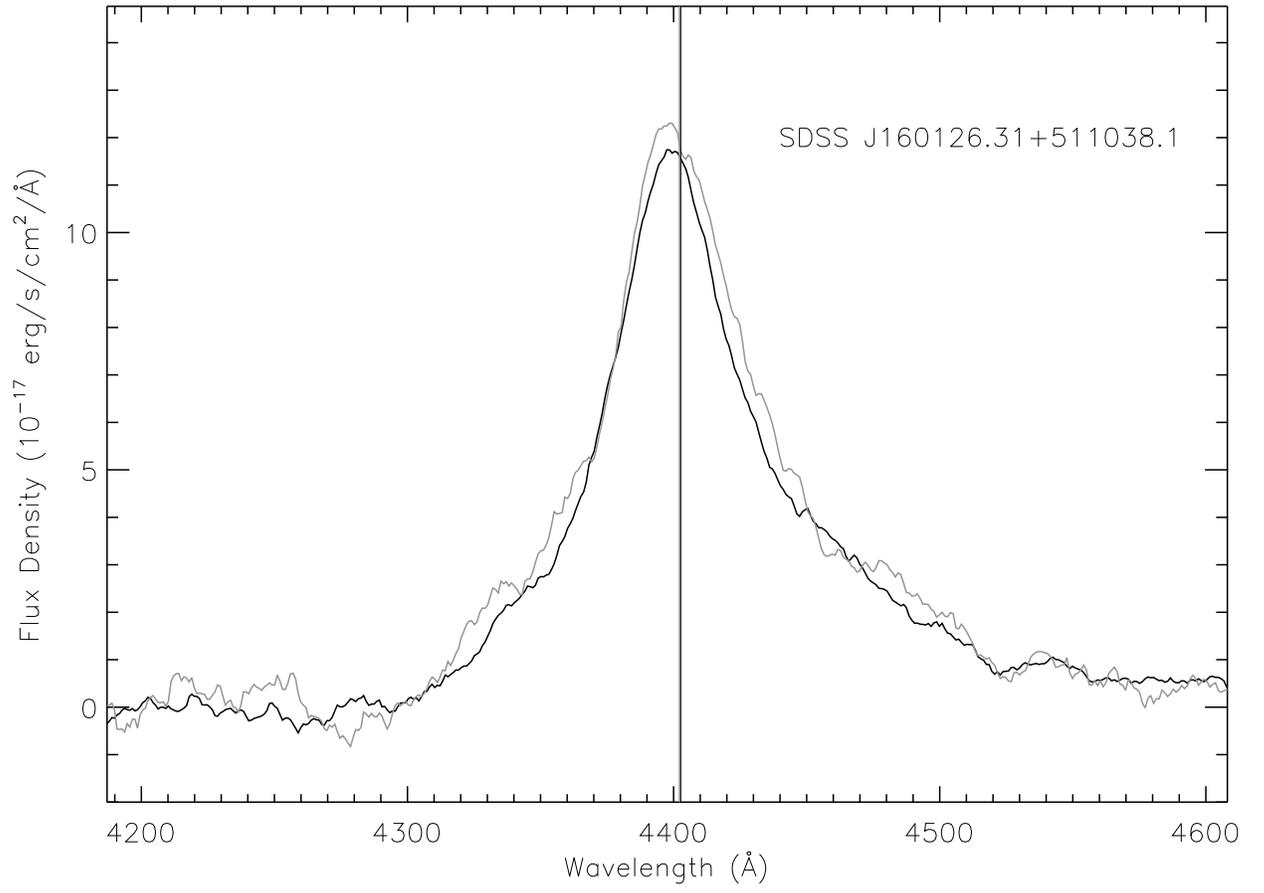}
\caption{C~\textsc{iv} line for quasar SDSS J231147.90+002941.9.  
\label{SDSSJ231147.90+002941.9fig}}
\end{figure}

SDSS J231147.90+002941.9 (Fig. \ref{SDSSJ231147.90+002941.9fig}) has line flux ratio $\frac{f_{HSN}}{f_{LSN}} = 1.03$ and width ratio $\frac{\sigma_{HSN}}{\sigma_{LSN}} = 1.03$ 

%and skewness difference $Pskew_{HSN} - Pskew_{LSN} = -0.53$.
%This large value for the change in skewness may be an overestimate, but there is certainly an increase in flux on the blue wing (and evidence for a possible decrease in the red wing).  

\begin{figure}
\includegraphics[scale=1.0]{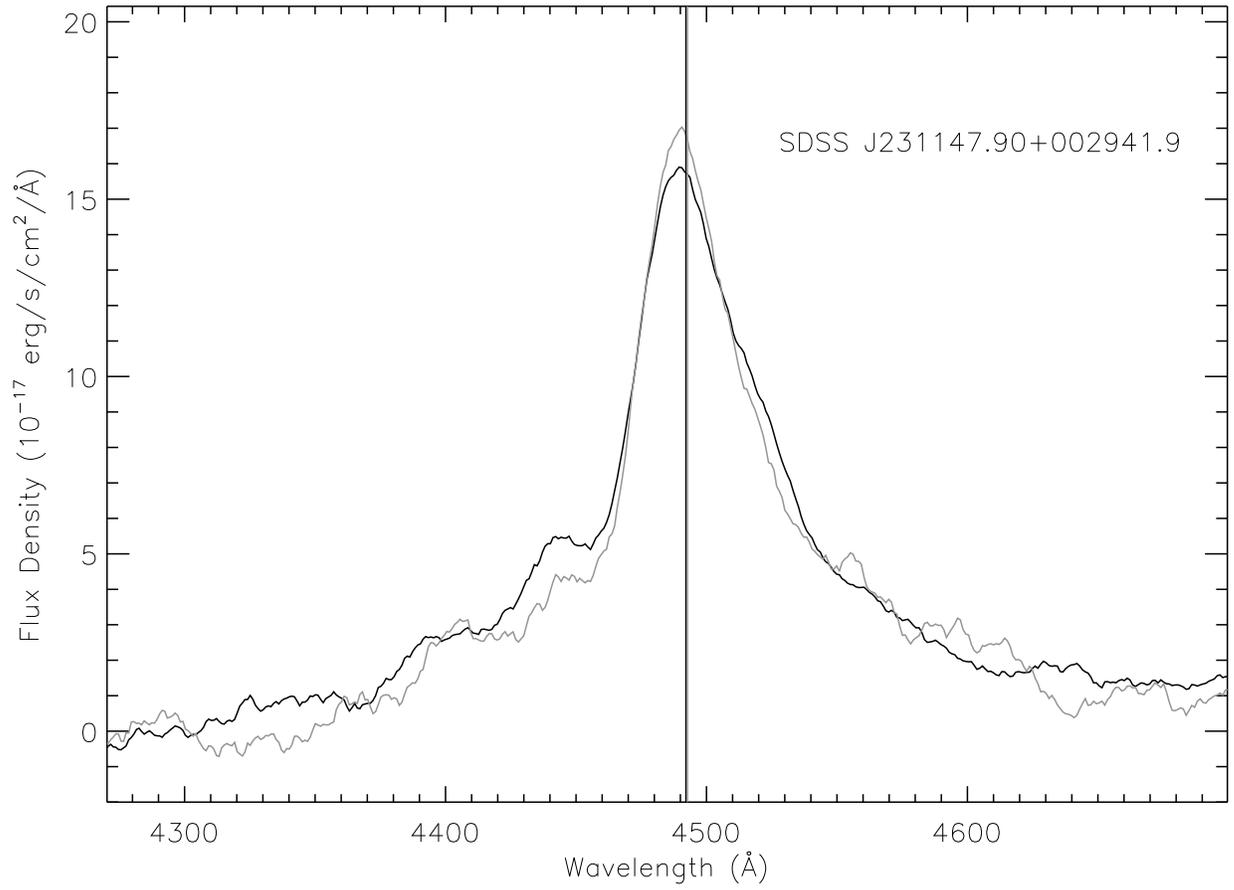}
\caption{C~\textsc{iv} line for quasar SDSS J160126.31+511038.1.  
\label{SDSSJ160126.31+511038.1fig}}
\end{figure}

SDSS J160126.31+511038.1 (Fig. \ref{SDSSJ160126.31+511038.1fig}) has line flux ratio $\frac{f_{HSN}}{f_{LSN}} = 0.92$ and width ratio $\frac{\sigma_{HSN}}{\sigma_{LSN}} = 0.97$.

%Again, much of the flux difference is a result of the increased width of this line at the low-S/N epoch.

%\subsection{$\sigma_{HSN}/\sigma_{LSN}$\label{sigmarat}}

%\subsection{$\Delta{Pskew}$\label{dpskew}}

\subsection{SDSS J115154.83$-$005904.6\label{SDSSJ115154.83-005904.6}}

\begin{figure}
\includegraphics[scale=1.0]{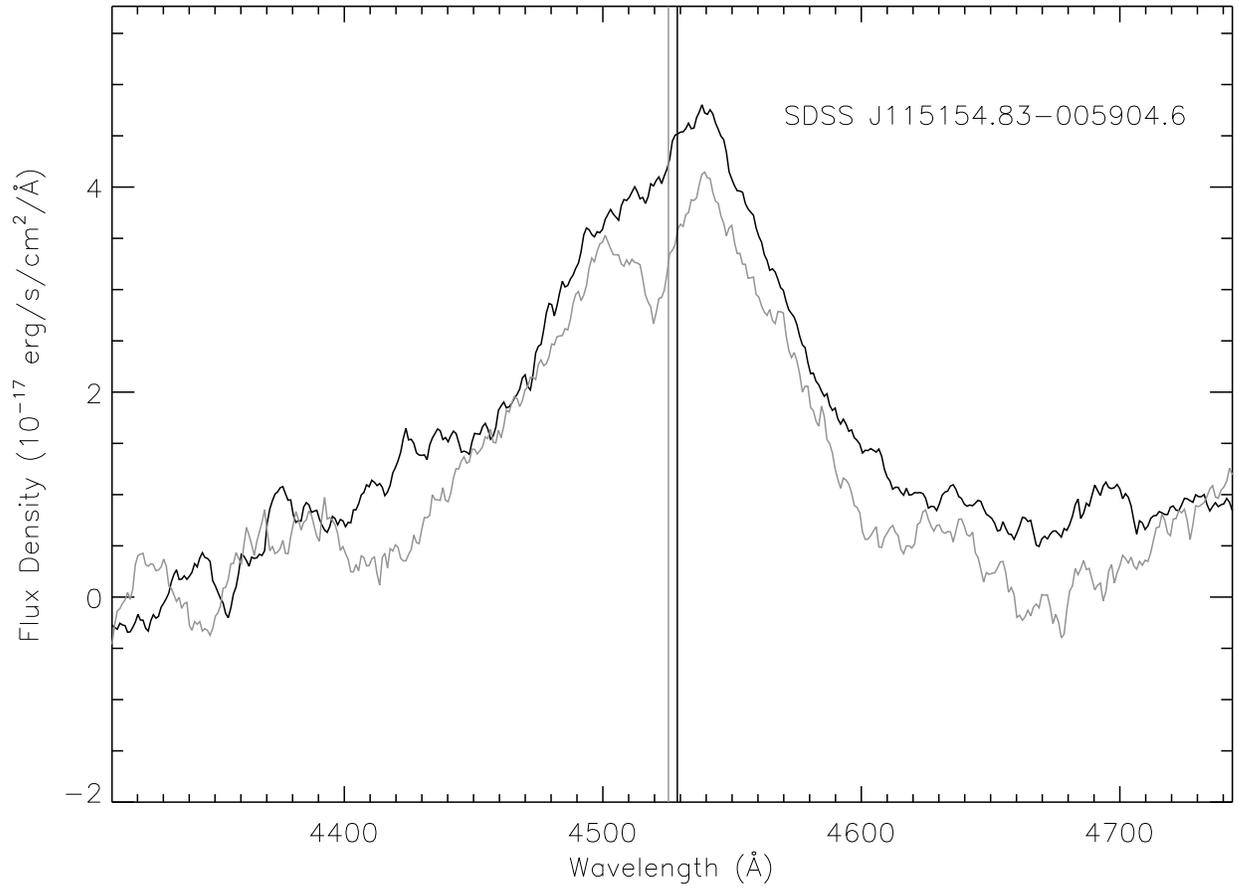}
\caption{C~\textsc{iv} line for quasar SDSS J115154.83$-$005904.6.  The bifurcation in the line is likely due to highly variable absorption.
\label{SDSSJ115154.83-005904.6fig}}
\end{figure}

Fig. \ref{SDSSJ115154.83-005904.6fig} shows the C~\textsc{iv} line of SDSS J115154.83$-$005904.6, a quasar at redshift $z = 1.93$.  This C~\textsc{iv} line was noticed as a part of the manual inspection for night sky and absorption lines done before line fitting.  The low-S/N epoch line (light curve) appears bifurcated, while the high-S/N epoch line (dark curve) does not; it is thus unlikely to be the result of an intervening absorption system.  These spectra were taken 281 days apart, corresponding to roughly 96 days' separation in the quasar rest-frame.  The depression near the peak of the line at the low-S/N epoch lies at $\sim4500$\AA\ in the observed frame, where there is not expected to be a large contamination from night sky lines.  This dip in flux near 4500\AA\ is not seen in other objects observed simultaneously with the same plate.  If this bifurcation is a real effect, this could potentially be an intriguing object for follow-up study.  The line profile looks similar to the C~\textsc{iv} line of NGC 3516, a well-studied intrinsic absorption system which varied by roughly 50\% in absorption equivalent width \citep{voit87}.  However, in the case of NGC 3516, the absorption was visible in all 11 epochs.  It should also be noted that NGC 3783, a nearby ($z=0.0097$) type I AGN has shown strong variations in many of its absorption features \citep{kraemer01,crenshaw03, gabel03}.  SDSS J115154.83$-$005904.6 shows little or no sign of absorption in the high-S/N epoch spectrum.

%%%%%%%%%%%%%%%%%%%%%%%%%%%%%%%%
%
% Discussion
%
%%%%%%%%%%%%%%%%%%%%%%%%%%%%%%%%

\section{Discussion\label{discussion}}

\subsection{Broad Line Region Structure\label{blrstructure}}

\citet{wills93} found that the equivalent width of the C~\textsc{iv} line decreased with increasing line width and that C~\textsc{iv} lines were typically asymmetric, skewed blueward of the flux peak.  They proposed the so-called ILR model to describe the structure of the broad-line region (BLR), in which the C~\textsc{iv} line flux is produced in two distinct portions of the BLR.  In this scheme, the narrow core of the C~\textsc{iv} line is produced in the "intermediate width emission line region" (ILR), located far from the quasar central engine.  The considerably broader base of the line arises in the very broad line region (VBLR), thought to be the portion of the BLR close to the central continuum source.  The line base is blueshifted with respect to the line core, presumably due to some relative bulk motion towards the observer by the VBLR gas; this blueshift of the line base produces a composite line which is skewed towards the blue.  From quasar to quasar, the range of possible equivalent widths for the narrow line core appears to be much larger than the range of possible equivalent widths for the broad line base.  This results in the overall line width depending primarily on the relative importance of the line core. In high-flux lines, the line core is dominant, resulting in high equivalent widths and narrow lines.  In low-flux lines, the line core does not dominate, the line base is relatively more important, and the line is broad.
This yields the single-epoch relation between equivalent width and line FWHM seen in Fig. 3 of \citet{wills93}.

Disk-wind models of the BLR, such as that of \citet{murray97}, are able to reproduce the C~\textsc{iv} line profiles first seen in \citet{wills93}, although they do not produce enough flux in the blue tail of the line. In such models, the line core and base are produced in different portions of one continuous BLR, rather than two distinct regions. The line base photons are simply emitted at a smaller distance from the central source, where the gas velocities are greater.  Instead of coming from just two distinct line production regions, the overall line profile can be thought of as the superposition of infinitely many lines lines whose width depends on the distance from the central black hole at which they were produced.

The intrinsic relation between C~\textsc{iv} flux and width---as an individual line gets stronger, it also becomes wider---may be explained through similar geometric arguments.  In fact, one should expect individual lines to broaden with increasing flux if one believes a) the broad portion of the line is produced near to the central engine and b) the portion of the line produced near to the central engine is more variable.  

That the line base is produced nearer to the central source is widely believed.  It is assumed that the increased width comes with the higher rotational velocities of the BLR gas nearer to the central black hole and thus, in a deeper potential well.

That the flux from the inner region of the BLR should be more variable is an essentially geometric argument.  The relatively small size allows for more coherent variability \citep{korista04}.  The larger the region, the more "washed out" the fluctuations in the continuum flux will be, due to a larger range in light-travel times from the various portions of the BLR to the observer.  This effect has been seen in reverberation mapping studies; recent attempts at reverberation mapping in high-luminosity objects have not yet produced conclusive results, due to the relative lack of variability in the emission lines \citep{kaspi03}.  The large ionizing flux in these high-luminosity systems pushes the BLR to great distances from the central engine, resulting in larger ranges in light-travel times throughout the BLR and decreased coherent variability.

Most current models make both of the assumptions necessary to explain the observed intrinsic flux-width relation.  Both assumptions are a part of the ILR and disk-wind models.  While this relation does not currently have the power to differentiate between discrete and continuous models of the BLR, it does strongly rule out any models which cannot predict such a relation.

It is also worth noting that in the \citet{wills93} ILR model, the C~\textsc{iv} line base is blueshifted with respect to the line core.  This suggests that the line skewness and median should also change with flux; as the more variable, blueshifted broad line base increases in relative importance, the median should be "dragged" to the blue and the line should be skewed bluewards as well.  That this is not seen at all in Figs. \ref{df.dz} and \ref{df.dpskew} is intriguing and merits further study.

\subsection{Orientation\label{orientation}}

\citet{richards02a} suggested that the size of the C~\textsc{iv}-Mg~\textsc{ii} line shift could be a function of the orientation of the quasar with respect to the observer or of the disk-wind opening angle.  
They noted that the C~\textsc{iv} blueshift seems to be the result of the degradation of the red side of the C~\textsc{iv} line, rather than a systemic shift of the entire emission line.  

A wide distribution of $\Delta{v}$ values (as seen in the lower panel of Fig. \ref{lineshift2panel}) suggests that individual blueshift measurements may not be reliable.  Even if the C~\textsc{iv}-Mg~\textsc{ii} blueshift is due to orientation, the measurements of the shift must be non-variable and reproducible in order to be a useful measure of viewing angle for an individual object, though the ensemble average would still be useful.  Fig. \ref{lineshift2panel} shows a large scatter in $\Delta{v}$ when the entire sample is included, casting doubt on the reliability of any single measurement of blueshift.  However, for the highest flux C~\textsc{iv} lines, we have demonstrated that the line shifts for these objects appear to be robust (see Fig. \ref{lineshift2panelhighflux}).  While this certainly does not constitute proof that the blueshift is an orientation effect, the opposite result would have caused serious problems for its use as a measure of orientation.  It is certainly possible to imagine a scenario where the C~\textsc{iv}-Mg~\textsc{ii} line shift is a product of viewing angle but varies in size.  However, in such a system, the line shift would not be useful as an orientation measure, even if it is an orientation effect.

Although the distribution of line shift differences between epochs, $\Delta{v}$, for the whole sample is impractically large, the narrow distribution for the highest flux objects indicates that the Mg~\textsc{ii}-C~\textsc{iv} line shift may hold promise as an orientation measure.  Observers are warned, however, that an adequate line flux is necessary to reliably measure the line shift.

\subsection{Line Flux Variability\label{discussionfluxvariability}}

Reverberation mapping studies indicate that the distance the BLR lies from the central source increases with continuum luminosity \citep[e.g.][]{wandel99,kaspi05}.  More luminous central sources produce a greater number of ionizing photons, "pushing" the BLR to greater distances.  As mentioned in \S\,\ref{blrstructure}, one would expect the emission lines from more luminous quasars to be less variable, as their larger sizes allow less coherent variability, "smearing out" the variations in the incident continuum over a greater range in light-travel times.  
The median Intrinsic Baldwin Effect Slope for our sample ($b_{int} = 0.15 \pm 0.06$) is shallower than that of the value of $b_{int} \approx 0.35$ from \citet{kinney90}.  This is to be expected, given that objects in our sample are much higher luminosity ($<M_{r}>=-25.6$) than those studied in \citet{kinney90}.  NGC 5548, for example, has an absolute magnitude $M_{R} \approx -22.5$ \citep{tyson98}.  

It is not clear, however, why there is no sign of a decrease in C~\textsc{iv} line flux variability with increasing luminosity (decreasing absolute magnitude) in Fig. \ref{civbothmr}.  If more luminous sources are less variable, and line variability is a result of continuum variations, one might expect line variability to share the anti-correlation with continuum luminosity.   It could simply be a function of binning; although our sample spans four magnitudes in luminosity, the difference between the median luminosities of the two most extreme bins is only 1.5 magnitudes.  However, as this was only a two-epoch study, and we have no knowledge, for individual objects, of the time lag between continuum and CIV variability, the issue may not be that simple.  A larger dynamic range, or a larger number of objects per bin, might yield a more illuminating result.

%%%%%%%%%%%%%%%
% Conclusions %
%%%%%%%%%%%%%%%

\section{Conclusions\label{conclusions}}

Using a sample of 105 quasars observed multiple times by the Sloan Digital Sky Survey, we have studied the variability of the C~\textsc{iv} line.  Spectra were fit using moment analysis techniques and four main conclusions are drawn:

1)  We find a strong correlation between the change in C~\textsc{iv} line flux and the change in line width.  As an individual quasar's C~\textsc{iv} line flux increases, so does the C~\textsc{iv} line width.  This is consistent with any picture of the BLR in which the broad line base is produced nearer to the central engine and the portion of the BLR nearer to the central engine exhibits more coherent line flux variability.

2)  We demonstrate that there is no apparent variability in the the blueshift of the C~\textsc{iv} line with respect to the Mg~\textsc{ii} line for the highest flux C~\textsc{iv} lines, a possibly positive sign for the use of line shifts as an orientation measure.

3)  With our measurements of continuum and line fluxes, we are able to reproduce the Baldwin Effect, deriving a slope of $b=0.78$.  We also calculate a median slope for the Intrinsic Baldwin Effect of $b_{int}=0.15$, shallower than the $b_{int} \approx 0.35$ determined by \citet{kinney90} for lower luminosity AGN.

4)  Using the continuum flux at the position of the line center, we reproduce well-known dependences of continuum variability amplitude on quasar luminosity and rest-frame time lag.  However, these same dependences are not evident for the amplitude of the C~\textsc{iv} line variability.  This may be due to the "smearing out" of continuum variability by the extended BLR.

Funding for the creation and distribution of the SDSS Archive has been provided by the Alfred P. Sloan Foundation, the Participating Institutions, the National Aeronautics and Space Administration, the National Science Foundation, the U.S. Department of Energy, the Japanese Monbukagakusho, and the Max Planck Society. The SDSS Web site is {\tt http://www.sdss.org/}.

The SDSS is managed by the Astrophysical Research Consortium (ARC) for the Participating Institutions. The Participating Institutions are The University of Chicago, Fermilab, the Institute for Advanced Study, the Japan Participation Group, The Johns Hopkins University, the Korean Scientist Group, Los Alamos National Laboratory, the Max-Planck-Institute for Astronomy (MPIA), the Max-Planck-Institute for Astrophysics (MPA), New Mexico State University, University of Pittsburgh, University of Portsmouth, Princeton University, the United States Naval Observatory, and the University of Washington.

%%%%%%%%%%%%%%
% References %
%%%%%%%%%%%%%%

\onecolumn
\clearpage
%%%%%%%%%%%
% Figures %
%%%%%%%%%%%
%
% use scale = 0.5, 0.3, 0.3 respectively

% \begin{figure}
% %%\epsscale{0.5}
% %%\plotone{image.ps}
% \includegraphics{f1.eps}
% \caption{ 
% \label{image1}}
% \end{figure}

%Section 2 Figures

%\begin{figure}
%\includegraphics[scale=1.0]{f1.eps}
%\caption{Difference in measured redshift between epochs as a function of high-S/N epoch line flux.  Each panel corresponds to a different method (described in \S\,\ref{linefitting}) used to measure the line center.  
%\label{dz4panel}}
%\end{figure}
%\clearpage

\clearpage

\clearpage

%\begin{figure}
%\includegraphics[scale=1.0]{f4.eps}
%\caption{C~\textsc{iv} line width versus line flux at the high-S/N epoch.  This is a reproduction of Figure 3 of \citet{wills93}.
%\label{f.sigma}}
%\end{figure}
%\clearpage

\clearpage

\clearpage

\clearpage

\clearpage

\clearpage

\clearpage

\clearpage

%\begin{figure}
%\includegraphics[scale=1.0]{f12.eps}
%\caption{Change in C~\textsc{iv} line width versus line flux change for objects with $f_{line,HSN} > 800 %\times 10^{-17}$ erg/s/cm$^{2}$.  Objects plotted with filled squares are discussed in \S\,\ref{individual}.
%\label{df.dsigma.high}}
%\end{figure}
%\clearpage

\clearpage
\clearpage

%\begin{figure}
%\includegraphics[scale=1.0]{f13.eps}
%\caption{C~\textsc{iv} line for quasar SDSS J082328.61 +061146.0.  
%\label{SDSSJ082328.61+061146.0fig}}
%\end{figure}
%\clearpage

\clearpage

\clearpage

%\begin{figure}
%\includegraphics[scale=1.0]{SDSSJ081931.48+055523.6.ps}
%\caption{C~\textsc{iv} line for quasar SDSSJ081931.48+055523.6.  
%\label{SDSSJ081931.48+055523.6fig}}
%\end{figure}
%\clearpage

\clearpage

%Tables

%%\clearpage
%%Table 1. -- Plate List
\begin{deluxetable}{clcccrccrr}
\rotate
\tablewidth{0pt}
%\setstretch{1.15}
\tableheadfrac{0}
\tablecaption{CIV Variability Sample.  HSN and LSN indicate the high- and low-S/N ratio epochs, respectively.  Redshifts and magnitudes are from the high-S/N ratio epoch. 
\label{civqsotable}}
\tablehead{
	\colhead{Number} &
	\colhead{SDSS J} & 
	\multicolumn{2}{c}{MJD} & 
	\colhead{$z_{HSN}$} & 
	\colhead{$\Delta{\tau}$} & 
  \colhead{$r_{HSN}$} &

	\colhead{$M_{r,HSN}$} & 
	\multicolumn{2}{c}{$S/N_{r}$}\\
	\colhead{} &
	\colhead{} &
	\colhead{HSN} &
	\colhead{LSN} &
	\colhead{} &
	\colhead{(days)} &
	\colhead{} &
	\colhead{} &
	\colhead{HSN} &
	\colhead{LSN}
}
\tablecolumns{10}
\startdata

  1 &        100013.37+011203.2 & 51910 & 51581 &  1.80 & 117.4 &  19.1 &  $-26.0$ &  12.3 &  11.4 \\

  2 &        100428.43+001825.6 & 51910 & 51581 &  3.04 &  81.3 &  18.7 &  $-27.6$ &  16.4 &  11.2 \\

  3 &      114211.59$-$005344.2 & 51959 & 51584 &  1.92 & 128.5 &  20.1 &  $-26.8$ &  22.7 &  16.4 \\

  4 &        114948.81+000855.8 & 51959 & 51584 &  1.97 & 126.3 &  19.4 &  $-26.2$ &  12.9 &  10.9 \\

  5 &      115154.83$-$005904.6 & 51943 & 51662 &  1.93 &  96.0 &  18.3 &  $-26.0$ &  12.9 &   7.6 \\

  6 &      115043.87$-$002354.0 & 51943 & 51662 &  1.98 &  94.4 &  19.7 &  $-27.8$ &  40.0 &  27.2 \\

  7 &        115213.55+001946.7 & 51943 & 51662 &  1.83 &  99.2 &  19.6 &  $-25.3$ &   8.0 &   7.3 \\

  8 &      124524.59$-$000937.9 & 51928 & 51660 &  2.08 &  86.9 &  19.5 &  $-27.3$ &  29.7 &  20.8 \\

  9 &      124356.22$-$000021.8 & 51928 & 51660 &  1.84 &  94.5 &  19.4 &  $-25.5$ &   9.8 &   7.2 \\

 10 &        124242.11+001157.9 & 51928 & 51660 &  2.16 &  84.8 &  18.4 &  $-26.3$ &  14.2 &   9.4 \\

 11 &      125617.52$-$001918.2 & 51689 & 51994 &  1.77 & 110.1 &  18.1 &  $-25.1$ &   7.9 &   5.5 \\

 12 &      125532.24$-$010608.7 & 51689 & 51994 &  1.78 & 109.5 &  18.1 &  $-25.2$ &   8.5 &  10.3 \\

 13 &        131630.46+005125.5 & 51985 & 51585 &  2.40 & 117.5 &  17.5 &  $-27.2$ &  18.3 &  11.7 \\

 14 &        131840.95+003103.9 & 51984 & 51665 &  1.77 & 115.0 &  18.9 &  $-25.3$ &   8.2 &   6.8 \\

 15 &        132214.82+005419.9 & 51959 & 51663 &  2.15 &  94.1 &  18.9 &  $-26.8$ &  19.6 &  15.0 \\

 16 &        133939.01+001021.6 & 51955 & 51662 &  2.13 &  93.7 &  20.3 &  $-26.1$ &  10.4 &   4.6 \\

 17 &      140114.28$-$004537.1 & 51942 & 51641 &  2.52 &  85.5 &  18.8 &  $-26.7$ &  14.2 &  12.8 \\

 18 &      135844.57$-$011055.1 & 51942 & 51641 &  1.96 & 101.6 &  19.1 &  $-26.3$ &  16.7 &  18.0 \\

 19 &      135605.41$-$010024.4 & 51942 & 51641 &  1.89 & 104.3 &  19.2 &  $-25.8$ &  12.6 &   8.9 \\

 20 &      135247.96$-$002351.6 & 51942 & 51641 &  1.67 & 112.7 &  18.8 &  $-24.7$ &   6.3 &   4.5 \\

 21 &        135828.74+005811.5 & 51942 & 51641 &  3.92 &  61.1 &  19.4 &  $-27.1$ &   9.5 &   7.1 \\

 22 &      142205.10$-$000120.7 & 51609 & 51957 &  1.86 & 121.7 &  19.4 &  $-26.0$ &  13.4 &  11.2 \\

 23 &        142209.11+005436.3 & 51609 & 51957 &  3.68 &  74.3 &  17.5 &  $-26.9$ &   8.3 &   9.1 \\

 24 &      145555.00$-$003713.4 & 51994 & 51666 &  1.95 & 111.3 &  19.1 &  $-25.4$ &   9.3 &   5.5 \\

 25 &      145302.09$-$010524.4 & 51994 & 51666 &  1.81 & 116.8 &  19.3 &  $-26.9$ &  27.6 &  17.9 \\

 26 &        145246.52+003450.5 & 51994 & 51666 &  2.54 &  92.6 &  19.8 &  $-24.6$ &   2.8 &   4.0 \\

 27 &        145429.65+004121.2 & 51994 & 51666 &  2.66 &  89.6 &  19.5 &  $-26.7$ &  14.1 &   9.7 \\

 28 &      131728.74$-$024759.4 & 51691 & 51990 &  3.38 &  68.3 &  17.8 &  $-26.7$ &   8.5 &   9.1 \\

 29 &        172909.93+624519.7 & 51694 & 51789 &  1.75 &  34.6 &  19.5 &  $-25.8$ &  10.6 &   8.2 \\

 30 &        022534.09+000347.9 & 51817 & 52238 &  1.74 & 153.9 &  18.8 &  $-24.2$ &   4.0 &   2.8 \\

 31 &      022346.42$-$003908.2 & 51817 & 52238 &  1.67 & 157.4 &  18.4 &  $-25.6$ &  11.9 &  15.3 \\

 32 &        022526.15+010124.0 & 51817 & 52238 &  1.87 & 146.5 &  18.4 &  $-24.9$ &   6.3 &   3.7 \\

 33 &      025038.68$-$004739.1 & 51816 & 51877 &  1.84 &  21.5 &  18.2 &  $-26.0$ &  15.4 &  17.3 \\

 34 &        025701.94+010644.6 & 51816 & 51877 &  2.19 &  19.1 &  19.6 &  $-24.5$ &   2.4 &   3.3 \\

 35 &        030600.41+010145.4 & 51817 & 51873 &  2.19 &  17.5 &  19.2 &  $-26.7$ &  14.6 &  10.8 \\

 36 &        031127.55+005357.3 & 51931 & 52254 &  1.76 & 117.1 &  19.9 &  $-25.5$ &  11.4 &   8.9 \\

 37 &      032253.09$-$001121.6 & 51929 & 51821 &  1.88 &  37.5 &  19.2 &  $-24.2$ &   2.9 &   3.8 \\

 38 &        031544.54+004220.9 & 51929 & 51821 &  1.88 &  37.5 &  18.8 &  $-25.6$ &  10.4 &   7.8 \\

 39 &      034318.37$-$004447.9 & 51811 & 51885 &  1.75 &  26.9 &  18.3 &  $-23.7$ &   2.9 &   1.5 \\

 40 &        003732.61+144258.0 & 51817 & 51884 &  2.37 &  19.9 &  19.4 &  $-24.5$ &   3.0 &   2.3 \\

 41 &        003520.91+143730.2 & 51817 & 51884 &  1.86 &  23.4 &  18.6 &  $-24.5$ &   4.8 &   3.2 \\

 42 &        003240.57+143951.9 & 51817 & 51884 &  1.86 &  23.4 &  19.9 &  $-24.3$ &   4.0 &   3.0 \\

 43 &        004337.73+160530.0 & 51868 & 51812 &  1.97 &  18.8 &  19.8 &  $-24.6$ &   3.5 &   3.7 \\

 44 &      093622.06$-$004555.4 & 52314 & 52027 &  1.78 & 103.4 &  18.7 &  $-24.3$ &   3.3 &   2.7 \\

 45 &      093736.74$-$000732.1 & 52314 & 52027 &  1.79 & 102.9 &  19.7 &  $-25.3$ &   8.3 &   5.1 \\

 46 &      093233.65$-$003441.9 & 52314 & 52027 &  1.84 & 101.2 &  19.8 &  $-24.7$ &   4.1 &   2.8 \\

 47 &      093150.57$-$001935.2 & 52314 & 52027 &  1.84 & 101.1 &  17.6 &  $-25.6$ &   8.8 &   8.3 \\

 48 &        094149.60+003254.3 & 52314 & 52027 &  2.00 &  95.6 &  18.2 &  $-24.4$ &   3.1 &   3.5 \\

 49 &        131522.44+013917.0 & 52295 & 52029 &  1.67 &  99.8 &  18.2 &  $-24.1$ &   3.5 &   3.1 \\

 50 &        131439.23+021214.9 & 52295 & 52029 &  1.78 &  95.8 &  20.3 &  $-24.8$ &   6.3 &   5.7 \\

 51 &        130754.44+021820.2 & 52295 & 52029 &  1.87 &  92.7 &  19.0 &  $-25.2$ &   7.9 &   4.4 \\

 52 &        130855.25+030614.2 & 52295 & 52029 &  1.79 &  95.5 &  19.3 &  $-25.3$ &   9.8 &   7.7 \\

 53 &        130940.60+031826.7 & 52295 & 52029 &  2.76 &  70.8 &  19.1 &  $-25.7$ &   5.8 &   3.3 \\

 54 &        130825.64+025736.0 & 52295 & 52029 &  1.75 &  96.6 &  19.0 &  $-25.5$ &  10.9 &   4.2 \\

 55 &        081859.78+423327.6 & 52207 & 51959 &  2.22 &  77.0 &  19.4 &  $-24.4$ &   3.7 &   2.3 \\

 56 &        081349.01+441517.7 & 52207 & 51959 &  2.21 &  77.1 &  18.6 &  $-26.0$ &  13.6 &  11.4 \\

 57 &        081614.97+435640.2 & 52207 & 51959 &  1.96 &  83.8 &  18.2 &  $-24.5$ &   5.2 &   4.8 \\

 58 &        081926.51+445759.9 & 52207 & 51959 &  1.71 &  91.4 &  19.8 &  $-25.0$ &   9.3 &   6.8 \\

 59 &        082310.94+442048.1 & 52207 & 51959 &  1.78 &  89.2 &  21.6 &  $-25.5$ &  13.9 &  12.6 \\

 60 &        153001.69+540452.3 & 52374 & 52442 &  1.72 &  25.0 &  19.0 &  $-25.9$ &  13.8 &   8.7 \\

 61 &        160126.31+511038.1 & 52375 & 52081 &  1.84 & 103.4 &  18.8 &  $-25.4$ &   9.4 &   3.7 \\

 62 &      231040.97$-$010823.0 & 52884 & 52534 &  2.07 & 114.2 &  19.3 &  $-25.8$ &   8.7 &   7.2 \\

 63 &      230952.29$-$003138.9 & 52884 & 52534 &  3.97 &  70.4 &  19.7 &  $-26.8$ &   6.7 &   4.5 \\

 64 &      230728.90$-$011608.9 & 52884 & 52534 &  1.98 & 117.4 &  19.6 &  $-25.1$ &   6.4 &   5.6 \\

 65 &      230832.98$-$002332.4 & 52884 & 52534 &  3.08 &  85.7 &  18.6 &  $-25.1$ &   2.6 &   1.1 \\

 66 &      230437.65$-$005703.3 & 52884 & 52534 &  2.49 & 100.2 &  19.5 &  $-25.5$ &   5.8 &   4.5 \\

 67 &      230402.78$-$003855.4 & 52884 & 52534 &  2.77 &  92.8 &  17.8 &  $-25.3$ &   3.3 &   2.4 \\

 68 &      230424.87$-$010140.8 & 52884 & 52534 &  1.89 & 121.1 &  19.8 &  $-24.2$ &   3.0 &   2.6 \\

 69 &        230239.68+002702.5 & 52884 & 52534 &  1.86 & 122.2 &  19.4 &  $-24.2$ &   3.1 &   3.1 \\

 70 &        230524.47+005209.7 & 52884 & 52534 &  1.85 & 123.0 &  18.9 &  $-25.1$ &   6.5 &   2.7 \\

 71 &        230323.77+001615.1 & 52884 & 52534 &  3.69 &  74.6 &  18.2 &  $-26.1$ &   4.3 &   3.6 \\

 72 &        230435.93+003001.5 & 52884 & 52534 &  2.00 & 116.6 &  19.2 &  $-24.6$ &   3.4 &   2.8 \\

 73 &        231121.98+004959.7 & 52884 & 52534 &  2.06 & 114.3 &  19.5 &  $-25.8$ &  10.4 &   5.6 \\

 74 &        231147.90+002941.9 & 52884 & 52534 &  1.90 & 120.6 &  19.4 &  $-25.5$ &   9.5 &   7.1 \\

 75 &        231241.77+002450.3 & 52884 & 52534 &  1.89 & 121.0 &  18.2 &  $-26.2$ &  15.9 &   7.7 \\

 76 &        224005.09+143147.8 & 52520 & 52264 &  3.49 &  57.0 &  19.9 &  $-26.1$ &   4.8 &   1.1 \\

 77 &        075153.67+331319.8 & 52237 & 52577 &  1.93 & 116.1 &  20.9 &  $-26.1$ &  13.9 &   7.6 \\

 78 &        075217.23+335524.5 & 52237 & 52577 &  1.68 & 126.7 &  19.0 &  $-24.8$ &   5.8 &   5.1 \\

 79 &        074823.86+332051.2 & 52237 & 52577 &  2.99 &  85.2 &  19.0 &  $-26.2$ &   6.3 &   9.1 \\

 80 &        075132.75+350535.0 & 52237 & 52577 &  2.07 & 110.9 &  19.4 &  $-24.7$ &   4.3 &   2.8 \\

 81 &        075321.93+350733.5 & 52237 & 52577 &  1.90 & 117.4 &  17.3 &  $-24.5$ &   3.8 &   5.0 \\

 82 &        144059.16+573724.3 & 52346 & 52433 &  2.04 &  28.6 &  20.0 &  $-25.5$ &   5.9 &   5.7 \\

 83 &        143618.60+581044.2 & 52346 & 52433 &  1.65 &  32.8 &  19.1 &  $-24.1$ &   3.9 &   2.3 \\

 84 &        145316.61+560750.8 & 52347 & 52435 &  1.85 &  30.9 &  18.2 &  $-24.2$ &   2.7 &   2.5 \\

 85 &        143632.31+563319.5 & 52347 & 52435 &  1.77 &  31.8 &  18.8 &  $-25.6$ &   8.9 &   7.0 \\

 86 &        161240.98+435749.4 & 52443 & 52355 &  1.74 &  32.1 &  19.0 &  $-24.3$ &   4.0 &   0.7 \\

 87 &        101902.02+473714.5 & 52347 & 52674 &  2.95 &  82.8 &  19.4 &  $-26.9$ &   9.7 &   7.8 \\

 88 &        102048.82+483908.8 & 52347 & 52674 &  1.94 & 111.2 &  20.2 &  $-26.5$ &  17.5 &  16.9 \\

 89 &        105922.46+494918.2 & 52346 & 52669 &  1.68 & 120.5 &  18.8 &  $-24.5$ &   2.7 &   5.9 \\

 90 &        105430.08+491947.1 & 52346 & 52669 &  4.00 &  64.6 &  17.7 &  $-27.1$ &   4.8 &   8.5 \\

 91 &        105027.74+490453.0 & 52346 & 52669 &  1.86 & 112.8 &  20.6 &  $-25.6$ &   6.2 &  11.1 \\

 92 &        104951.09+493156.2 & 52346 & 52669 &  1.79 & 115.7 &  19.5 &  $-25.3$ &   7.4 &   9.6 \\

 93 &        104806.47+501021.5 & 52346 & 52669 &  1.78 & 116.0 &  19.7 &  $-24.2$ &   3.8 &   3.6 \\

 94 &        105454.16+503123.9 & 52346 & 52669 &  1.87 & 112.4 &  19.1 &  $-26.3$ &  19.1 &  12.4 \\

 95 &        074641.95+293247.9 & 52346 & 52663 &  2.28 &  96.8 &  18.8 &  $-26.2$ &  11.8 &  10.0 \\

 96 &        074407.41+294707.4 & 52346 & 52663 &  1.86 & 110.8 &  19.0 &  $-25.8$ &  10.9 &  14.0 \\

 97 &        074625.28+302020.7 & 52346 & 52663 &  1.74 & 115.9 &  19.4 &  $-26.6$ &  27.8 &  21.7 \\

 98 &        074937.74+304021.4 & 52346 & 52663 &  1.73 & 116.2 &  19.2 &  $-24.1$ &   3.7 &   4.8 \\

 99 &        082443.39+055503.7 & 52962 & 52737 &  2.10 &  72.5 &  19.4 &  $-26.5$ &  13.2 &   9.0 \\

100 &        082328.61+061146.0 & 52962 & 52737 &  2.78 &  59.5 &  19.1 &  $-27.9$ &  24.9 &  14.6 \\

101 &        082256.01+060528.7 & 52962 & 52737 &  1.98 &  75.4 &  19.7 &  $-25.7$ &   9.4 &   4.2 \\

102 &        081941.12+054942.6 & 52962 & 52737 &  1.70 &  83.3 &  19.8 &  $-24.0$ &   2.8 &   2.3 \\

103 &        081931.48+055523.6 & 52962 & 52737 &  1.69 &  83.7 &  19.9 &  $-26.6$ &  21.1 &  12.6 \\

104 &        081811.50+053713.9 & 52962 & 52737 &  2.51 &  64.1 &  19.0 &  $-27.2$ &  17.7 &  11.2 \\

105 &        082257.04+070104.3 & 52962 & 52737 &  2.95 &  56.9 &  18.2 &  $-27.5$ &  17.5 &  10.2 \\
\enddata
\end{deluxetable}

\clearpage
\begin{deluxetable}{ccccc}
%\rotate

\tableheadfrac{0}
\tablewidth{0pt}
\tablecaption{CIV Flux Variability. \label{civfluxtable}}
\tablehead{
	\colhead{Number} &
	\multicolumn{2}{c}{$f_{cont}$} &
	\multicolumn{2}{c}{$f_{line}$} \\
	\colhead{} &
	\multicolumn{2}{c}{($10^{-17}$ erg/s/cm$^{2}$/\AA)} & 
	\multicolumn{2}{c}{($10^{-17}$ erg/s/cm$^{2}$)} \\
	\colhead{} &
	\colhead{HSN} &
	\colhead{LSN} &
	\colhead{HSN} &
	\colhead{LSN}
}
\tablecolumns{5}
\startdata

  1 &   11.2 &   20.3 & 1205.1 & 1099.3 \\

  2 &    8.0 &    6.4 & 1248.3 & 1165.5 \\

  3 &   19.6 &   27.5 & 1458.3 & 1644.1 \\

  4 &    9.7 &   13.5 &  746.4 &  654.4 \\

  5 &   10.1 &    7.1 &  614.6 &  479.6 \\

  6 &   50.3 &   44.9 & 4691.5 & 3729.1 \\

  7 &    5.8 &    8.7 &  712.1 &  853.0 \\

  8 &   20.7 &   17.3 & 1991.0 & 2013.4 \\

  9 &    7.6 &    4.8 &  760.7 &  702.2 \\

 10 &    8.9 &    6.3 &  784.8 &  925.1 \\

 11 &    5.9 &    3.3 &  514.7 &  513.1 \\

 12 &    4.3 &    6.5 &  374.1 &  484.6 \\

 13 &   12.0 &    9.9 & 1943.9 & 1937.0 \\

 14 &    4.5 &    6.4 &  771.7 &  780.5 \\

 15 &   12.9 &   15.0 &  918.3 &  971.4 \\

 16 &    6.3 &    3.6 &  722.0 &  864.0 \\

 17 &    6.7 &    8.2 &  496.0 &  657.7 \\

 18 &   10.3 &   14.3 &  600.1 &  624.7 \\

 19 &    8.1 &    4.5 &  386.1 &  386.2 \\

 20 &    4.5 &    3.0 &  423.3 &  517.9 \\

 21 &    2.6 &    2.1 &  406.8 &  306.3 \\

 22 &   10.2 &    8.3 &  958.0 &  917.8 \\

 23 &    2.6 &    1.5 &  109.5 &  822.0 \\

 24 &    5.3 &    3.6 &  408.2 &  493.3 \\

 25 &   24.1 &   19.4 & 2014.2 & 1861.1 \\

 26 &    0.7 &    1.9 &  200.4 &  269.8 \\

 27 &    5.4 &    4.5 &  898.1 &  736.5 \\

 28 &    2.7 &    3.3 &  655.0 &  591.7 \\

 29 &    8.1 &    7.0 & 1335.1 & 1290.9 \\

 30 &    2.3 &    1.3 &  466.1 &  489.1 \\

 31 &    8.5 &   14.5 &  359.0 &  374.4 \\

 32 &    4.1 &    2.4 &  311.5 &  297.3 \\

 33 &    9.3 &   14.9 &  748.2 & 1071.2 \\

 34 &    1.1 &    1.3 &   81.2 &  160.4 \\

 35 &   10.1 &    8.0 &  926.0 &  717.4 \\

 36 &    6.6 &    4.9 &  341.4 &  432.5 \\

 37 &    2.0 &    2.1 &  365.1 &  388.5 \\

 38 &    6.5 &    4.9 &  483.7 &  492.2 \\

 39 &    1.0 &    0.1 &  390.5 &  244.4 \\

 40 &    1.2 &    0.9 &  220.6 &  209.5 \\

 41 &    2.3 &    1.8 &  395.6 &  331.1 \\

 42 &    1.8 &    1.4 &  326.1 &  304.9 \\

 43 &    2.2 &    2.3 &  374.8 &  451.4 \\

 44 &    2.4 &    2.0 &  330.4 &  242.5 \\

 45 &    4.2 &    3.2 &  302.9 &  241.1 \\

 46 &    3.1 &    2.2 &  384.0 &  319.5 \\

 47 &    5.8 &    8.1 & 1217.8 & 1203.8 \\

 48 &    1.2 &    1.6 &  294.4 &  409.3 \\

 49 &    2.3 &    3.2 &  173.6 &  169.1 \\

 50 &    3.8 &    5.7 &  451.6 &  457.0 \\

 51 &    4.2 &    2.5 &  437.8 &  437.6 \\

 52 &    5.1 &    6.7 &  584.1 &  735.5 \\

 53 &    2.0 &    1.4 &   77.7 &   44.9 \\

 54 &    8.3 &    4.1 &  411.7 &  352.2 \\

 55 &    1.1 &    0.9 &   81.8 &   76.6 \\

 56 &    5.0 &    6.5 &  507.5 &  458.2 \\

 57 &    2.4 &    2.6 &  213.4 &  219.6 \\

 58 &    4.4 &    3.4 &  499.1 &  457.6 \\

 59 &    7.0 &    8.7 &  769.1 &  857.1 \\

 60 &   12.5 &   10.3 &  572.4 &  493.5 \\

 61 &    6.5 &    4.1 &  958.7 & 1038.7 \\

 62 &    4.7 &    6.6 &  459.6 &  560.6 \\

 63 &    2.0 &    2.3 &  218.3 &  297.8 \\

 64 &    2.7 &    3.9 &  120.5 &  355.3 \\

 65 &    0.9 &    0.8 &   20.0 &   27.6 \\

 66 &    2.4 &    3.0 &  275.3 &  260.5 \\

 67 &    1.0 &    1.9 &  359.2 &  380.5 \\

 68 &    1.8 &    1.7 &  278.1 &  357.3 \\

 69 &    2.3 &    2.6 &  274.3 &  329.3 \\

 70 &    4.2 &    2.2 &  452.6 &  631.8 \\

 71 &    0.9 &    1.1 &  368.5 &  332.8 \\

 72 &    1.7 &    2.8 &  377.8 &  312.5 \\

 73 &    6.3 &    5.3 &  445.3 &  326.8 \\

 74 &    6.9 &    8.8 & 1452.4 & 1403.6 \\

 75 &   11.0 &    6.9 &  689.3 &  572.7 \\

 76 &    1.5 &    0.8 &  275.1 &  329.0 \\

 77 &    8.0 &    4.1 &  417.1 &  377.8 \\

 78 &    5.4 &    3.5 &  413.2 &  456.2 \\

 79 &    2.4 &    3.5 &  427.7 &  254.7 \\

 80 &    1.7 &    2.4 &  173.6 &  167.5 \\

 81 &    2.2 &    3.1 &  249.6 &  262.4 \\

 82 &    4.2 &    5.5 &  541.7 &  547.4 \\

 83 &    2.6 &    2.7 &  356.9 &  157.1 \\

 84 &    2.0 &    0.9 &  507.8 &  507.9 \\

 85 &    7.4 &    3.6 &  764.6 &  604.3 \\

 86 &    3.0 &    2.7 &  395.5 &  351.4 \\

 87 &    5.3 &    3.7 &  405.5 &  433.5 \\

 88 &   15.6 &   18.9 & 1497.7 & 1537.6 \\

 89 &    4.1 &    5.2 &  483.2 &  698.9 \\

 90 &    2.0 &    2.5 &  441.3 &  790.4 \\

 91 &    7.1 &   10.8 &  927.3 &  973.0 \\

 92 &    6.3 &    8.1 &  796.7 &  775.6 \\

 93 &    2.0 &    3.4 &  555.5 &  421.5 \\

 94 &   14.8 &   11.1 & 1135.4 & 1212.0 \\

 95 &    6.6 &    5.2 &   38.5 &   78.8 \\

 96 &    7.2 &    9.7 &  501.5 &  421.5 \\

 97 &   21.8 &   16.6 & 3051.4 & 2626.4 \\

 98 &    1.4 &    2.2 &  245.2 &  218.1 \\

 99 &   10.7 &    8.4 &  898.4 &  863.8 \\

100 &   16.3 &   13.8 & 1383.6 & 1104.5 \\

101 &    6.3 &    4.3 &  456.6 &  412.7 \\

102 &    2.1 &    3.9 &  786.5 &  602.5 \\

103 &   28.0 &   15.2 & 1447.8 & 1044.2 \\

104 &   11.3 &   13.6 &  884.4 &  930.1 \\

105 &    8.9 &    6.8 &  786.6 &  917.3 \\
\enddata
\end{deluxetable}

\clearpage
\begin{deluxetable}{ccccccc}
%\rotate

\tableheadfrac{0}
\tablewidth{0pt}
\tablecaption{CIV Profile Variability. \label{thesisprofiletable}}
\tablehead{
	\colhead{Number} &
	\multicolumn{2}{c}{$\lambda_{median}$} &
	\multicolumn{2}{c}{$\sigma_{CIV}$} &
	\multicolumn{2}{c}{$Pskew$} \\
	\colhead{} &
	\multicolumn{2}{c}{(\AA)} & 
	\multicolumn{2}{c}{(\AA)} & 
	\colhead{} &
	\colhead{} \\
	\colhead{} &
	\colhead{HSN} &
	\colhead{LSN} &
	\colhead{HSN} &
	\colhead{LSN} &
	\colhead{HSN} &
	\colhead{LSN}
}
\tablecolumns{7}
\startdata

  1 & 4335.96 & 4336.67 &  51.8 &  50.3 & -0.02 & -0.04 \\

  2 & 6254.13 & 6255.22 &  60.2 &  55.8 & -0.16 & -0.14 \\

  3 & 4504.86 & 4505.81 &  54.3 &  54.0 & -0.00 & -0.03 \\

  4 & 4587.67 & 4591.58 &  55.5 &  45.8 & -0.03 & -0.07 \\

  5 & 4528.83 & 4525.40 &  60.7 &  54.8 & -0.08 & -0.19 \\

  6 & 4609.86 & 4610.34 &  51.4 &  49.1 & -0.11 & -0.10 \\

  7 & 4386.77 & 4384.37 &  50.6 &  53.3 & -0.15 & -0.07 \\

  8 & 4755.69 & 4755.93 &  60.9 &  61.7 &  0.05 &  0.04 \\

  9 & 4386.39 & 4389.90 &  47.8 &  47.8 & -0.13 & -0.13 \\

 10 & 4885.17 & 4884.34 &  46.9 &  55.8 & -0.04 & -0.04 \\

 11 & 4283.01 & 4280.57 &  41.5 &  46.1 &  0.03 & -0.08 \\

 12 & 4320.61 & 4321.66 &  53.9 &  53.3 &  0.02 & -0.01 \\

 13 & 5267.12 & 5266.23 &  55.5 &  54.9 & -0.07 & -0.10 \\

 14 & 4292.50 & 4294.13 &  56.6 &  50.0 & -0.03 &  0.05 \\

 15 & 4861.69 & 4864.76 &  66.6 &  60.0 & -0.04 & -0.03 \\

 16 & 4836.85 & 4836.54 &  47.4 &  59.8 & -0.01 & -0.09 \\

 17 & 5436.41 & 5440.46 &  68.1 &  71.6 &  0.10 &  0.10 \\

 18 & 4577.88 & 4567.58 &  60.2 &  55.9 & -0.12 &  0.22 \\

 19 & 4473.25 & 4472.83 &  52.3 &  52.4 & -0.11 &  0.04 \\

 20 & 4132.72 & 4133.57 &  46.2 &  40.6 & -0.04 & -0.00 \\

 21 & 7611.62 & 7608.21 & 109.3 &  97.5 &  0.09 &  0.14 \\

 22 & 4426.41 & 4428.08 &  46.5 &  46.6 & -0.09 & -0.10 \\

 23 & 7187.91 & 7215.40 &  54.9 & 127.7 &  0.28 &  0.19 \\

 24 & 4560.30 & 4558.58 &  49.8 &  55.2 & -0.01 &  0.00 \\

 25 & 4339.19 & 4337.84 &  51.0 &  51.2 &  0.02 &  0.05 \\

 26 & 5473.89 & 5483.46 &  72.3 &  73.9 & -0.11 & -0.10 \\

 27 & 5651.77 & 5652.11 &  62.4 &  49.1 &  0.07 &  0.03 \\

 28 & 6796.05 & 6797.81 &  88.2 &  81.0 & -0.34 & -0.35 \\

 29 & 4252.94 & 4251.77 &  43.1 &  42.7 &  0.14 &  0.11 \\

 30 & 4229.30 & 4228.77 &  49.7 &  50.4 & -0.07 & -0.11 \\

 31 & 4128.10 & 4125.95 &  44.5 &  54.5 &  0.06 &  0.15 \\

 32 & 4446.01 & 4445.38 &  46.6 &  52.5 & -0.03 & -0.18 \\

 33 & 4390.24 & 4392.37 &  59.3 &  53.7 & -0.03 & -0.10 \\

 34 & 4917.46 & 4926.01 &   \nodata &  80.8 &   \nodata &  0.00 \\

 35 & 4932.10 & 4934.35 &  60.9 &  59.4 & -0.01 & -0.07 \\

 36 & 4270.86 & 4269.77 &  50.9 &  54.6 & -0.21 & -0.20 \\

 37 & 4458.80 & 4457.78 &  53.4 &  54.9 & -0.02 & -0.12 \\

 38 & 4451.02 & 4454.29 &  46.2 &  49.3 &  0.06 & -0.03 \\

 39 & 4251.56 & 4253.38 &  56.1 &  48.3 & -0.08 & -0.05 \\

 40 & 5200.92 & 5204.93 &  65.9 &  62.9 &  0.07 &  0.16 \\

 41 & 4424.24 & 4425.90 &  58.2 &  53.1 & -0.16 & -0.25 \\

 42 & 4428.53 & 4430.25 &  53.4 &  54.4 & -0.00 & -0.13 \\

 43 & 4605.88 & 4604.88 &  50.5 &  58.3 &  0.07 &  0.00 \\

 44 & 4295.65 & 4295.44 &  42.3 &  35.6 & -0.14 &  0.51 \\

 45 & 4313.70 & 4314.42 &  56.0 &  61.3 &  0.01 & -0.03 \\

 46 & 4383.73 & 4390.22 &  53.6 &  46.4 & -0.05 &  0.06 \\

 47 & 4390.96 & 4391.03 &  48.9 &  42.9 &  0.03 &  0.09 \\

 48 & 4661.90 & 4663.15 &  54.9 &  48.5 & -0.21 & -0.01 \\

 49 & 4108.36 & 4134.69 &  53.2 &  45.9 &  0.31 & -0.31 \\

 50 & 4294.97 & 4290.71 &  47.3 &  49.7 &  0.03 &  0.05 \\

 51 & 4434.07 & 4434.83 &  54.3 &  56.1 &  0.10 &  0.17 \\

 52 & 4305.35 & 4306.14 &  53.5 &  51.3 &  0.09 &  0.11 \\

 53 & 5818.30 & 5748.08 &  80.1 & 177.3 &  0.07 & -2.35 \\

 54 & 4259.49 & 4261.48 &  50.5 &  39.0 &  0.17 &  0.35 \\

 55 & 4992.64 & 4986.64 &  63.4 &  57.8 & -0.14 & -0.22 \\

 56 & 4960.77 & 4958.50 &  62.3 &  59.2 &  0.15 &  0.20 \\

 57 & 4581.82 & 4574.54 &  59.2 &  60.9 & -0.01 & -0.07 \\

 58 & 4189.89 & 4187.46 &  48.4 &  52.4 &  0.04 & -0.00 \\

 59 & 4301.59 & 4302.12 &  48.2 &  48.8 &  0.01 & -0.05 \\

 60 & 4203.21 & 4201.92 &  62.3 &  63.4 &  0.09 &  0.19 \\

 61 & 4402.60 & 4402.21 &  43.5 &  44.6 &  0.13 &  0.05 \\

 62 & 4727.93 & 4733.90 &  62.4 &  68.4 &  0.09 &  0.22 \\

 63 & 7668.57 & 7671.34 &  66.2 &  64.7 & -0.11 & -0.41 \\

 64 & 4633.27 & 4631.65 &   \nodata &  63.9 &   \nodata & -0.09 \\

 65 & 6341.47 & 6312.69 &   \nodata &   \nodata &   \nodata &   \nodata \\

 66 & 5407.90 & 5412.07 &  69.6 &  54.1 & -0.14 & -0.30 \\

 67 & 5836.82 & 5838.44 &  73.2 &  64.3 & -0.21 & -0.13 \\

 68 & 4482.85 & 4482.89 &  47.2 &  57.0 & -0.35 & -0.08 \\

 69 & 4428.81 & 4423.38 &  50.5 &  57.6 &  0.04 &  0.10 \\

 70 & 4404.43 & 4403.24 &  53.9 &  61.9 &  0.05 &  0.06 \\

 71 & 7269.31 & 7266.72 &  79.8 &  89.4 & -0.13 & -0.14 \\

 72 & 4651.40 & 4650.14 &  64.1 &  52.9 & -0.13 &  0.03 \\

 73 & 4719.64 & 4720.61 &  64.7 &  56.9 &  0.11 &  0.11 \\

 74 & 4492.13 & 4492.69 &  54.4 &  52.9 & -0.05 &  0.01 \\

 75 & 4480.50 & 4482.09 &  52.3 &  52.9 &  0.15 & -0.04 \\

 76 & 6964.24 & 6969.20 &  82.1 &  95.3 & -0.14 & -0.36 \\

 77 & 4542.01 & 4541.62 &  48.4 &  45.9 & -0.07 & -0.05 \\

 78 & 4147.37 & 4149.49 &  41.7 &  46.3 & -0.12 & -0.06 \\

 79 & 6148.37 & 6148.65 &  85.2 &  73.4 &  0.28 &  0.45 \\

 80 & 4738.29 & 4745.81 &  63.6 &  64.6 &  0.09 &  0.02 \\

 81 & 4488.77 & 4490.48 &  54.2 &  47.5 & -0.17 & -0.10 \\

 82 & 4688.73 & 4694.29 &  64.8 &  57.7 &  0.14 &  0.04 \\

 83 & 4105.50 & 4104.59 &  53.8 &   \nodata & -0.10 &   \nodata \\

 84 & 4402.43 & 4400.79 &  40.4 &  44.1 &  0.00 & -0.11 \\

 85 & 4269.64 & 4273.86 &  57.8 &  63.2 &  0.06 & -0.00 \\

 86 & 4243.85 & 4244.72 &  61.3 &  58.9 &  0.12 & -0.16 \\

 87 & 6087.18 & 6091.37 &  70.0 &  81.6 &  0.19 &  0.02 \\

 88 & 4546.91 & 4547.56 &  47.3 &  51.2 &  0.05 &  0.06 \\

 89 & 4141.26 & 4150.70 &  41.2 &  47.3 & -0.02 & -0.22 \\

 90 & 7723.83 & 7725.32 &   \nodata &  73.8 &   \nodata &  0.05 \\

 91 & 4431.05 & 4431.47 &  47.7 &  48.1 &  0.04 &  0.08 \\

 92 & 4316.52 & 4314.17 &  53.0 &  49.1 & -0.05 & -0.19 \\

 93 & 4301.31 & 4306.88 &  49.4 &  41.4 &  0.09 &  0.33 \\

 94 & 4449.11 & 4449.00 &  52.3 &  56.6 & -0.02 & -0.05 \\

 95 & 5050.78 & 5056.25 &   \nodata &  59.1 &   \nodata & -0.63 \\

 96 & 4418.00 & 4419.19 &  61.1 &  54.0 & -0.06 & -0.02 \\

 97 & 4229.98 & 4229.32 &  48.8 &  48.8 & -0.06 & -0.05 \\

 98 & 4204.25 & 4216.29 &  56.7 &  64.4 &  0.34 &  0.03 \\

 99 & 4788.89 & 4787.13 &  47.6 &  53.2 & -0.08 & -0.10 \\

100 & 5855.95 & 5859.41 &  70.1 &  60.7 & -0.20 & -0.23 \\

101 & 4614.60 & 4614.45 &  45.4 &  52.1 & -0.02 &  0.01 \\

102 & 4172.34 & 4174.75 &  66.4 &  67.5 &  0.06 &  0.03 \\

103 & 4154.89 & 4153.73 &  46.4 &  41.5 & -0.21 & -0.11 \\

104 & 5437.31 & 5437.47 &  65.5 &  67.4 &  0.22 &  0.18 \\

105 & 6102.28 & 6105.53 &  73.5 &  77.5 &  0.03 & -0.04 \\
\enddata
\end{deluxetable}

\clearpage

%%
%% The End
%%
\end{document}